\documentclass[pdflatex,sn-mathphys]{sn-jnl_cln}

\usepackage{amsmath}
\usepackage{mathtools}

\usepackage{cleveref}
\usepackage{xcolor}
\usepackage{csvsimple}
\usepackage{subcaption}

\usepackage{siunitx}

\usepackage[utf8]{inputenc}
\usepackage[font={small}]{caption}
\usepackage{makecell, cellspace}%, caption}

\usepackage{graphicx}

\usepackage{booktabs}
\newcolumntype{H}{>{\setbox0=\hbox\bgroup}c<{\egroup}@{}}

\usepackage{tabularx}
\usepackage{xltabular}

\usepackage{algorithm,algorithmicx,algpseudocode}
\usepackage{multirow}
\usepackage{multicol}
\usepackage[numbers]{natbib}

\usepackage{perpage} %the perpage package
\MakePerPage{footnote} %the perpage package

\usepackage{array} % needed for below macros
\newcolumntype{C}[1]{>{\raggedleft\arraybackslash}p{#1}}
\newcolumntype{P}[1]{>{\raggedright\arraybackslash}p{#1}}

\newcommand{\ignore}[1]{}

\usepackage{xcolor}  % Add this line in the preamble if not already there

\newcommand{\setItemSep}{\setlength\itemsep}

\begin{document}

% suggested titles
% BERT-based Conflict Detection in SRS documents
% 
% \title{Adaptive Fine-tuning for Multiclass Requirement Classification}
\title[Optimizing Hard-to-Place Kidney Allocation]{\vspace{-70pt}
\fontsize{14pt}{18pt}\selectfont Optimizing Hard-to-Place Kidney Allocation: A Machine Learning Approach to Center Ranking \vspace{-10pt}}

\author{{\fontsize{9pt}{9pt}\selectfont Sean Berry \hfill} \\[-3pt]{\fontsize{7pt}{9pt}\selectfont Department of Mechanical, Industrial and Mechatronics Engineering, Toronto \hfill} \\[-5pt]
{\fontsize{7pt}{9pt}\selectfont Metropolitan University, Toronto, ON M5S 3G8, sean.berry@torontomu.ca \hfill} \\
{\fontsize{9pt}{9pt}\selectfont Berk G{\"o}rg{\"u}l{\"u} \hfill} \\[-3pt] {\fontsize{7pt}{9pt}\selectfont McMaster University, DeGroote School of Business, Hamilton, ON L8S 4M4, \hfill} \\[-5pt]
{\fontsize{7pt}{9pt}\selectfont gorgulub@mcmaster.ca \hfill} \\
{\fontsize{9pt}{9pt}\selectfont Sait Tun\c{c} \hfill} \\[-3pt] {\fontsize{7pt}{9pt}\selectfont Virginia Tech, Grado Department of Industrial and Systems Engineering, Blacksburg, \hfill} \\[-5pt]
{\fontsize{7pt}{9pt}\selectfont VA 24061, sait.tunc@vt.edu \hfill} \\
{\fontsize{9pt}{9pt}\selectfont Mucahit Cevik \hfill} \\[-3pt] {\fontsize{7pt}{9pt}\selectfont Department of Mechanical, Industrial and Mechatronics Engineering, Toronto \hfill} \\[-5pt]
{\fontsize{7pt}{9pt}\selectfont Metropolitan University, Toronto, ON M5S 3G8, mcevik@torontomu.ca \hfill} \\
{\fontsize{9pt}{9pt}\selectfont Matthew J. Ellis \hfill} \\[-3pt]{\fontsize{7pt}{9pt}\selectfont Duke University School of Medicine, Department of Medicine, Durham, NC 27710 \phantom{aaaaaa}}\\[-20pt]
}

\abstract{
Kidney transplantation is the preferred treatment for end-stage renal disease, yet the scarcity of donors and inefficiencies in allocation systems create major bottlenecks, resulting in prolonged wait times and alarming mortality rates. Despite their severe scarcity, timely and effective interventions to prevent non-utilization of life-saving organs remain inadequate. Expedited out-of-sequence placement of hard-to-place kidneys to centers with the highest likelihood of utilizing them has been recommended in the literature as an effective strategy to improve placement success. Nevertheless, current attempts towards this practice is non-standardized and heavily rely on the subjective judgment of the decision-makers. This paper proposes a novel data-driven, machine learning-based ranking system for allocating hard-to-place kidneys to centers with a higher likelihood of accepting and successfully transplanting them. Using the national deceased donor kidney offer and transplant datasets, we construct a unique dataset with donor-, center-, and patient-specific features. We propose a data-driven out-of-sequence placement policy that utilizes machine learning models to predict the acceptance probability of a given kidney by a set of transplant centers, ranking them accordingly based on their likelihood of acceptance. Our experiments demonstrate that the proposed policy can reduce the average number of centers considered before placement by fourfold for all kidneys and tenfold for hard-to-place kidneys. This significant reduction indicates that our method can improve the utilization of hard-to-place kidneys and accelerate their acceptance, ultimately reducing patient mortality and the risk of graft failure. Further, we utilize machine learning interpretability tools to provide insights into factors influencing the kidney allocation decisions.

% Our work demonstrate the substantial potential of machine learning to revolutionize the kidney allocation process. by

% We find that the proposed machine learning-based approach shows great promise in improving the kidney allocation process.

% As such, there is a possibility that our approach could contribute to enhancing transplantation rates. However, further research and validation are needed to fully understand the impact and to confirm if this could lead to an increased number of lives saved.
}

% \keywords{Software Requirement Specification (SRS), Conflict Detection, Semantic Similarity, Named Entity Recognition (NER)}
%%%%%%%%%%%%%%%%%%%%%%%%%%%%%%%%%%%%%%%%%%%%%%%%%%%%%%%%%%%%%%%%%%%%%%%%%%%%%%%%%

\maketitle

%%%%%%%%%%%%%%%%%%%%%%%%%%%%%%%%%%%%%%%%%%%%%%%%%%%%%%%%%%%%%%%%%%%%%%%%%%%%%%%%%
\section{Introduction}\label{sc:Intro}

%%%%% Motivation

% kidney disease
Kidney disease is the ninth leading cause of death in the US and represents a significant global health challenge, with millions of people affected by end-stage renal disease (ESRD) \citep{kochanek2024mortality}. The preferred and most effective treatment for ESRD is a kidney transplant. However, the scarcity of donor kidneys and inefficiencies in allocation systems create significant bottlenecks, leading to lengthy wait times and increased waitlist mortality rates \citep{gill2005impact, lewis2021organ}. This scarcity is further exacerbated by the high non-utilization rates of viable kidneys, which often result from a combination of logistical challenges, stringent organ selection criteria of transplant centers, and the time-sensitive nature of organ transplantation \citep{stewart2017diagnosing}. The non-utilization rate for hard-to-place kidneys, such as those from older donors, donors with medical comorbidities, or kidneys with extended ischemic times is particularly concerning. Non-utilization rates increase steadily with higher Kidney Donor Profile Index (KDPI) scores, a metric that combines several factors to estimate the risk of graft failure~\citep{bae2016changes}.

Despite the severe scarcity of life-saving organs, timely and effective measures to prevent non-utilization remain inadequate. One direct approach to increase the utilization of organs that might otherwise go unused is to impose penalties for their rejection. However, such penalties are prohibited in the U.S. Alternatively, hard-to-place organs can be offered to patients who have declared their willingness to accept such organs at the time of registration on the waiting list. This approach is partially implemented in the U.S., where patients can opt to be listed for organs with high Kidney Donor Profile Index (KDPI), hepatitis B and C virus (HCV+), and other specific conditions. Additionally, transplant centers can apply filters based on donor characteristics such as age or donation after cardiac death (DCD). Although the effectiveness of this practice has not been formally evaluated, the persistently high non-utilization rates highlight the urgent need for more effective interventions.

Expedited out-of-sequence placement of hard-to-place kidneys to transplant centers with the highest likelihood of utilization has been recommended in the literature as an effective strategy to rescue kidneys at high risk of non-utilization \citep{cooper2019report, reese2016new}. In line with this recommendation, the Organ Procurement and Transplantation Network (OPTN) implemented the Kidney Accelerated Placement (KAP) project from July 18, 2019, to July 15, 2020 \citep{mohan2022accelerating}. The KAP project aimed to increase the utilization of hard-to-place kidneys by offering them to centers with a proven history of accepting such organs. However, this initiative did not result in statistically significant improvements in the donor acceptance rate for hard-to-place kidneys compared to the year prior to the program start~\citep{noreen2022kidney}. Similarly, the U.K.'s Kidney Fast Track Scheme (KFTS) sought to allocate organs to the centers most likely to accept them after multiple rejections in the national allocation system \citep{callaghan2017early, white2015impact}. Unlike KAP, KFTS made simultaneous offers to all participating transplant centers, allowing the accepting center to transplant the kidney to any of their patients. Despite these efforts, KFTS also failed to significantly improve organ utilization. These outcomes highlight the need for more effective, data-driven interventions to enhance the allocation and utilization of hard-to-place kidneys.

In current practice, to increase the utilization and expedite the placement of hard-to-place kidneys, organ procurement organizations (OPOs) can deviate from the regular allocation process and extend out-of-sequence offers to transplant centers that are more likely to accept such kidneys \citep{adler2021greater}. However, this practice is currently non-standardized and relies heavily on the subjective judgment of individual OPOs and is often influenced by behind-the-scenes relationships, further raising concerns about transparency. The lack of defined guidelines for such allocation exceptions could potentially worsen disparities in organ access \citep{hanaway2020exacerbating}. Additionally, recent updates to the kidney allocation system have inadvertently created delays in local kidney placements, further increasing the prevalence of out-of-sequence offers. Consequently, there is an urgent need to leverage comprehensive data on center-level offer acceptance and transplant outcomes to develop an evidence-based approach to out-of-sequence offers. By doing so, we can ensure that expedited placement is trackable and is used judiciously, transparently, and only when necessary. Standardizing the criteria for bypassing the regular allocation process and making out-of-sequence offers would reduce reliance on ad hoc decisions and promote equity and efficiency in kidney allocation.

% operational inefficiencies
The complexities of kidney allocation are compounded by various factors, including the intricate decision-making process of accepting a donor kidney, the temporal dynamics of organ viability, geographic logistics, and the rapidly changing health status of patients. Specifically, transplant centers' acceptance decisions are shaped by factors such as donor-recipient compatibility, waiting time, the urgency of the recipient's condition, the distance between donor and recipient, and the center's infrastructure and expertise \citep{wolfe2007developing}. Additionally, transplant centers’ risk tolerance, which can fluctuate based on their recent outcomes, further complicates these decisions. Centers experiencing challenges with post-transplant outcomes may be more hesitant to accept higher-risk organs. These decisions can also vary depending on the individual risk tolerance of the on-call clinician, adding another layer of variability to the process. 

Due to the complexity of the data and the objectives, existing interventions to improve the placement likelihood of hard-to-place kidneys typically rely on simple heuristics and non-standardized criteria. However, these methods often fail to account for the dynamic and multifaceted nature of organ transplantation, potentially contributing to inherent discrepancies in organ allocation systems, such as disparities in transplant access, wait times, and waitlist mortality. Furthermore, a significant consequence of the reliance on simple heuristics is the high non-utilization rates of viable kidneys, as the system may struggle to find suitable recipients within the necessary timeframe. This issue underscores the need for more dynamic, data-driven methods to enhance the allocation process. By leveraging comprehensive data and advanced algorithms, we can develop more sophisticated models that better account for the complex factors influencing kidney placement, ultimately improving the efficiency and equity of organ allocation and reducing the non-utilization rates of viable kidneys.

Another key challenge in the kidney allocation and transplantation process is the overwhelming volume of data, which is difficult for humans to parse effectively or consistently. Simple heuristics and standardized criteria often fail to make efficient use of this data. In contrast, machine learning (ML) algorithms excel at analyzing large datasets, uncovering patterns that are indiscernible to humans. These algorithms can be employed to identify optimal donor-recipient matches, improving overall outcomes in kidney transplantation and reducing non-utilization rates.

Despite several studies investigating the kidney allocation problem, there remains a significant gap in optimizing the efficiency of organ allocation interventions to improve utilization. The current high non-utilization rates of donor kidneys, coupled with the growing demand for kidneys, underscore the urgent need for more effective intervention strategies. An ML-based ranking system can play a crucial role in addressing this challenge. Leveraging the predictive power of ML, we can better match donor organs with suitable recipients, thus reducing the likelihood of viable organs going unused. Such a system can maximize the utilization likelihood of hard-to-place kidneys by identifying and targeting transplant centers with the highest likelihood of acceptance and successful transplantation. By making more informed and data-driven interventions, this approach has the potential to significantly improve the efficiency and equity of the allocation of hard-to-place kidneys, ultimately saving more lives.

% % process improvement idea
% Overall, the kidney allocation process involves far too much data for humans to parse by hand, and using simple heuristics typically does not make efficient use of the data. ML algorithms can quickly and easily analyze large volumes of data, learning patterns indiscernible to humans. Hence, these algorithms can be used to identify donor-recipient matches, improving overall outcomes in kidney transplantation and lowering kidney non-utilization rates. While several studies have investigated the kidney allocation problem, there remains a significant gap in maximizing organ utilization and minimizing non-use.

% % need for intervention
% Given the inherent challenges in this problem, there is a significant need for an intervention that can help bridge the gap between the high non-utilization rates of donor kidneys and the growing need for kidneys in the American populace. By leveraging the predictive power of machine learning, we can better match donor organs with suitable recipients, thus reducing the likelihood of viable organs not being used. 
% % need for ranking 
% In addressing this challenge, a machine learning-based ranking system may prove to be helpful. 
% % This approach seems optimal for the problem, since to maximize utilization the system should aim to 
% Such an approach can maximize kidney utilization by making the best offers possible as it can help target donor kidneys to transplant centers with the highest likelihood of acceptance and successful transplantation.

%%%%% Research Goal
The objective of this research is to tackle the challenges in kidney allocation processes by developing a data-driven, ML-based approach. Our primary goal is to create an effective and interpretable ranking system for the allocation of hard-to-place kidneys to transplant centers. This approach leverages a comprehensive set of features, many of which have not been previously explored in the literature. Notably, we incorporate center-specific features such as the number of kidneys previously accepted by a center with higher Cold Ischemia Time (CIT) or Kidney Donor Risk Index (KDRI) than the offered kidney, as well as the proximity of transplant centers and donor hospitals to each other and to medium and large airports.

To showcase the effectiveness of our proposed framework, we compare its performance against the data from the current kidney allocation system as well as several heuristics that are inspired by the previously implemented interventions. Our primary evaluation criterion is the number of centers that reject an organ before it is ultimately accepted. Additionally, we provide a detailed analysis of both the local and global interpretability of the underlying prediction models. This analysis not only enhances clinicians' understanding of the ML model's inference process but also highlights the significance of center-specific features in improving kidney allocation outcomes. By offering a transparent and robust solution, our approach aims to contribute to more efficient and equitable organ allocation, ultimately reducing organ non-use and improving outcomes for patients awaiting transplants.

% The goal of this research is to address these challenges through a data-driven, machine-learning-based approach to kidney allocation. We aim to develop an effective and interpretable ranking system for the allocation of high-risk kidneys to centers. Our approach utilizes a wide array of features not previously considered in the literature, most prominently center-specific features such as the number of kidneys with higher cold ischemia times accepted, or the distance to medium and large airports. We then demonstrate the effectiveness of our method by comparing it against the ground truth number of rejections before an organ is accepted along with other heuristics. Furthermore, we offer a detailed analysis of both the local and global interpretability of the underlying prediction model, hence aiding clinicians' understanding of the model's decisions, as well as unravelling the importance of center-specific features in the allocation process.

Our research on optimizing the allocation of hard-to-place organs is both timely and highly relevant, given the urgent challenges in organ allocation and utilization. In response to these challenges, the OPTN Board of Directors established the Expeditious Task Force in September 2023~\citep{healio2024}, with the goal of increasing the number of transplanted organs by improving the efficiency of the organ placement process. The task force plans to actively engage the donation and transplant community, conducting multiple rapid, small-scale trials of innovative approaches to enhance organ usage and placement efficiency. Additionally, with the U.S. Health Resources and Services Administration (HRSA) currently seeking to revamp the OPTN system, our research comes at a pivotal moment. By introducing a novel data-driven approach to kidney transplantation interventions—ranking centers based on their likelihood of accepting hard-to-place organs, our work not only addresses existing inefficiencies but also aligns with ongoing policy discussions aimed at improving transplantation outcomes. This alignment underscores the potential impact of our research in shaping the future of organ allocation practices and contributing meaningfully to ongoing efforts to optimize the national transplant system.\\

%%%%% Contributions
%\subsection{Contributions}
We summarize the contributions of our work as follows:
\begin{itemize}\setItemSep{0.3em}
    \item We develop a novel ML-based ranking system for allocating hard-to-place kidneys to centers with a higher likelihood of accepting and successfully transplanting them. Our approach leverages a more comprehensive and diverse set of features than previously utilized in the literature, thereby significantly enhancing the predictive power and applicability of the proposed kidney allocation intervention framework.

    \item We introduce a robust framework for rigorously evaluating our model against existing heuristics and baseline data. This framework not only demonstrates the superior performance of our approach compared to current practices but also establishes a new benchmark for evaluating the performance of kidney allocation interventions more broadly.

    \item We conduct a comprehensive analysis of model explainability, addressing both local and global interpretability for the ML models used. This analysis increases the transparency of the allocation process of the hard-to-place organs and offers valuable insights that can inform the development of effective decision-support tools in transplantation. 
\end{itemize}

% Our work contributes to the literature on kidney allocation in several ways. First, we develop a machine learning (ML)-based ranking system to allocate kidney offers to centers. This ranking system is novel and not previously explored in the literature. Furthermore, our study distinguishes itself by utilizing a more extensive set of features in our machine learning models compared to previous attempts to address this problem.
% Next, we develop a novel framework for testing our model against other heuristics as well as the baseline data. This approach shows our model to be more effective than the current practice. In addition, we investigate the variation of feature importance when the Kidney Donor Risk Index (KDRI) is restricted to different ranges. Our study also highlights the importance of center-specific features for the prediction task.
% Finally, we explore the explainability of our model, providing an analysis of both local and global interpretability for this problem. This analysis should not only contribute to the transparency of the allocation process but also offer valuable insights for the development of decision-support tools, potentially building a stronger foundation for trust and collaboration between machine learning systems and medical professionals.

%%%%% Organization of the paper
The remainder of the paper is structured as follows. In Section~\ref{litrev}, we review the relevant literature to contextualize our research. Section~\ref{meth} outlines the comprehensive data and the overall methodology employed in this study. Section~\ref{res} presents and discusses the results of our analysis. Finally, Section~\ref{conc} concludes the paper by summarizing our findings and suggesting potential directions for future research.

%%%%%%%%%%%%%%%%%%%%%%%%%%%%%%%%%%%%%%%%%%%%%%%%%%%%%%%%%%%%%%%%%%%%%%%%%%%%%%%%%

%%%%%%%%%%%%%%%%%%%%%%%%%%%%%%%%%%%%%%%%%%%%%%%%%%%%%%%%%%%%%%%%%%%%%%%%%%%%%%%%%
\section{Literature Review}
\label{litrev}

The field of organ transplantation has seen significant growth in research, particularly in recent years, as efforts intensify to improve patient outcomes and optimize organ utilization. Despite its transformative potential, the application of ML models in this field remains relatively underexplored compared to traditional approaches~\citep{schold2023promise}. ML models may not reach their full potential if they are not applied correctly, either due to the lack of relevant and high-quality data needed to train these models effectively or because the problem at hand may not be well-suited to ML techniques~\citep{chui2018ai}. For instance, issues like overfitting in small datasets, the interpretability of complex models, or the mismatch between the model's assumptions and the nature of the medical data can hinder the effectiveness of ML solutions~\citep{lee2017medical}. A recent example is the study by \citet{truchot2023machine}, which compares various ML models with traditional Cox proportional hazard models. The study reveals that Cox models often perform equivalently or even better than ML models in predicting graft survival in kidney transplantation. These types of studies underscore the importance of careful consideration in how ML is integrated into transplantation medicine, emphasizing the need for further research to determine when and how ML can best complement or enhance existing methodologies.

One of the most pressing challenges in kidney transplantation is the high non-utilization rate of donor kidneys, particularly those classified as high-risk. \citet{crannell2022deceased} reported that nearly 45\% of deceased donor kidneys with a high KDPI were not used in the U.S. To alleviate this issue, several intervention mechanisms and accelerated placement schemes have been implemented globally, with varying degrees of success. The KAP project in the U.S. is a notable example, where transplant centers with a history of accepting similar organs were prioritized for offers (see \Cref{tab:kap} for details on the eligibility criteria used in KAP). Despite these efforts, systematic analyses have shown that such schemes have had limited impact, often due to their constrained scope and the subjective criteria used for allocation~\citep{noreen2022kidney, white2015impact}. Recent research has also attempted to predict the risk of non-use for donor kidneys. For instance, studies by \citet{massie2010improving} and \citet{marrero2017predictors} developed logistic regression models that outperformed KDRI in predicting non-use risk. Building on this, \citet{barah2021predicting} and \citet{li2024improving} proposed ML models, including random forests, boosting trees, and neural networks, which demonstrated strong predictive performance in identifying kidneys at risk of non-utilization.

\begin{table}[!ht]
\centering
\caption{Summary of donor characteristics and qualification thresholds used for eligibility in the KAP project.}
\label{tab:kap}

\begin{tabular}{P{0.4\textwidth} P{0.35\textwidth}}
\toprule 
\textbf{Donor Characteristic} & \textbf{Qualification Threshold} \\
\midrule
KDPI** & $\geq$ KDPI \\
Age (years) & $\geq$ Age - 10\% * Age \\
Peak serum creatinine (mg/dL) & $\geq$ Creatinine - 25\% * Creatinine \\
History of diabetes & Yes if yes, Yes or No if no \\
History of IV drug use & Yes if yes, Yes or No if no \\
DCD status & Yes if yes, Yes or No if no\\
\bottomrule
\end{tabular}

\end{table}

In the broader context of transplant medicine, ML has shown promise across various applications, including disease diagnosis, treatment response prediction, and patient management~\citep{gotlieb2022promise}. \citet{gotlieb2022promise} provide a comprehensive review of 36 publications, highlighting the potential of ML models while also emphasizing the need for robust validation, ethical considerations, and collaboration between clinicians, researchers, and data scientists. This underscores the importance of carefully integrating ML into clinical practice to ensure that its benefits are fully realized. \citet{connor2021future} provide a literature survey in several areas of interest for ML and organ transplantation namely clinical prediction and decision support, listing for transplantation, organ allocation, and prediction of patient and graft survival. They note key technical and ethical challenges that must be addressed for ML to become an effective clinical tool.

% \citet{gotlieb2022promise} provide a review of ML in transplant medicine, compiling the results of 36 different publications. The authors note that ML models have demonstrated promise in various aspects of transplantation medicine, including disease diagnosis, treatment response prediction, and patient management. The authors also note the need for robust validation of ML models, the ethical implications of ML applications, and the necessity for increased collaboration between clinicians, researchers, and data scientists.

% \citet{mark2019using} make use of a random survival forest model with conditional inference trees as base learners, for the purposes of predicting kidney transplant survival, which is shown to perform well for this task. The authors note that better survival predictions could eventually lead to more efficient allocation of kidneys and improve patient outcomes. Another interesting result is the suggestion of building separate models for different cohorts of patients, citing different top features for different cohorts. 

A notable gap in the existing literature is the use of CIT and center-specific, behavior-related features in predicting the acceptance or non-use of donor kidneys by transplant centers. The importance of CIT in transplantation outcomes cannot be overstated (see, for example, \citet{debout2015each}, demonstrating a proportional relationship between CIT and the risk of graft failure and patient survival). Advances in preservation techniques, such as improved preservation fluids or pumping, have mitigated but not eliminated the risks associated with prolonged CIT~\citep{ponticelli2015impact}. Therefore, omitting CIT as a predictive feature could lead to significant oversight, underestimating the complexities of organ acceptance and rejection decisions by transplant centers. Furthermore, the inclusion of center-specific features, such as the number of kidneys with higher CIT or KDPI previously accepted by a center, and logistical factors like proximity of transplant centers and donor hospitals to airports, has been largely overlooked in the literature. These variables are crucial for understanding how specific transplant centers are likely to respond to organ offers, especially for hard-to-place kidneys. By utilizing these features in the ML model training, a more nuanced and accurate prediction of organ acceptance can be achieved, ultimately improving kidney allocation efficiency.

The application of ML in kidney allocation holds tremendous potential for enhancing transplantation outcomes and reducing the rate of organ non-use. Our research stands out by focusing on optimizing interventions specifically for the allocation of hard-to-place kidneys---a topic that, to our knowledge, has not been thoroughly explored in the literature. We incorporate key features that are crucial in predicting the offer acceptance likelihood of transplant centers, including dynamic calculations of CIT, the historical acceptance behavior of centers, and various logistical factors. By addressing these critical gaps, our study aims to deliver a comprehensive, data-driven approach to kidney allocation that could significantly improve both the efficiency and equity of the transplant system.

% The application of ML in kidney allocation holds immense potential for improving transplantation outcomes and reducing non-use rates. 
% % While there is ongoing research in this field, it is important to further investigate the performance of these models. 
% % Furthermore, efforts should be directed towards allocation strategies for high-risk kidneys where the non-use rate is especially high.
% Furthermore, we distinguish the research by both prioritizing high-risk kidneys where the non-use rate is especially high, which are often overlooked in traditional allocation models, and through the dynamic calculation of cold ischemia time. Our study aims to fill these research gaps by providing a comprehensive data-driven approach to the kidney allocation problem.
%%%%%%%%%%%%%%%%%%%%%%%%%%%%%%%%%%%%%%%%%%%%%%%%%%%%%%%%%%%%%%%%%%%%%%%%%%%%%%%%%

%%%%%%%%%%%%%%%%%%%%%%%%%%%%%%%%%%%%%%%%%%%%%%%
\section{Methodology}
\label{meth}
%%%%%%%%%%%%%%%%%%%%%%%%%%%%%%%%%%%%%%%%%%%%%%%
This section provides a comprehensive overview of the dataset, experimental setup, modeling specifications, and interpretability methods employed in the paper.

%%%%%%%%%%%%%%%%%%%%%%%%%%%%%%%%%%%%%%%%%%%%%%%
\subsection{Dataset}
%%%%%%%%%%%%%%%%%%%%%%%%%%%%%%%%%%%%%%%%%%%%%%%
To analyze the likelihood of transplant centers accepting hard-to-place kidneys, we construct a unique dataset from two primary sources. First, we obtained deceased donor data from the OPTN, which includes records from 248,008 deceased donors between January 2016 and September 2021, all of whom had at least one kidney recovered for transplantation. The second source is the Potential Transplant Recipient (PTR) data for the same period, which documents all kidney offers made to patients on the US waiting list through a matching process called ``match run". We excluded donors missing key features, such as clamp data and initial response date, as well as kidney offers without a corresponding entry in the deceased donor dataset (further details are provided in \Cref{sc:data_exc}).
%\mycomment{S: Are we reporting all of these. If not, please create a Missing Data Imputation and Data Exclusion part (you don't need a seperate section, just use a sub-title) to report our approach and provide numbers (percent of data excluded).\textcolor{blue}{added.}}

% donors from the analysis due to missing match run creation times.

We create the initial dataset by merging donor-specific features from the deceased-donor dataset with kidney offers from the PTR data. During a match run, a kidney may be offered to multiple patients within the same transplant center. For each center, we consider only the first offer (to the patient with the highest priority at that center) or an accepted offer, if applicable. This approach enables us to assign a binary outcome (accept or reject) to each center-kidney pair. Additionally, we generate several center- and donor-specific variables, including the distance to large and medium airports, the historical acceptance rate of each center, and the number of kidneys with a higher KDRI accepted by the center in the past two years. 

The PTR data includes the offer times, which indicate when an organ is offered to a potential recipient's transplant center for consideration, for each kidney offer, allowing us to determine CIT---the period during which a donated organ is procured and remains outside the human body before transplantation. For our analysis, CIT is calculated as the difference between the clamp time (when blood supply to the organ is ceased and it is removed from the donor) and the offer time. \Cref{tbl:basic_stat_recipient_cat,tbl:basic_stat_recipient_cont} list the categorical and continuous features used in our analysis, respectively, along with summary statistics (e.g., distribution for categorical variables, mean/variance for continuous variables) stratified by the outcome.

\sisetup{round-mode=places, round-precision=1, table-format=2.2, table-number-alignment=center, round-pad=false}

\begin{table}[htbp]
\caption{Deceased donor kidney offers between 2016 and 2021 stratified by offer response: Categorical features.}
% \mycomment{S: Sean, Tables 1 and 2 are generated from Tables A1 and A2's content. Please carefully check to make sure that everything looks correct.} \mycomment{Also, please add counts for each category, under the first row, highlighted in red.} 
\label{tbl:basic_stat_recipient_cat}
\centering

\begin{minipage}{0.49\textwidth}
\resizebox{\linewidth}{!}{%
\begin{tabular}{llSS}
\toprule

&                           & \multicolumn{1}{c}{Accept} & \multicolumn{1}{c}{Reject} \\ \toprule
N                                                         &                 & \num{3944}    & \num{431125}    \\
\midrule
Donor Ethnicity & White  & \num{62.0}\% & \num{69.0}\% \\ 
                           & Black  & \num{22.0}\% & \num{16.0}\% \\
                           & Hispanic & \num{12.0}\% & \num{11.0}\% \\
                           & Other & \num{4.0}\% &  \num{4.0}\%\\
\midrule
Donor Gender    & M               & \num{52.0}\%    & \num{58.0}\%    \\
                & F               & \num{48.0}\%    & \num{42.0}\%    \\
\midrule
Donor Blood Type& O               & \num{46.0}\%    & \num{49.0}\%    \\
                & A               & \num{40.0}\%    & \num{40.0}\%    \\
                & B               & \num{14.0}\%    & \num{11.0}\%    \\
\midrule
Cause of Death  & CVD/Stroke & \num{52.0}\% & \num{45.0}\% \\
                & Anoxia          & \num{32.0}\%    & \num{41.0}\%    \\
                & Head Trauma     & \num{13.0}\%    & \num{11.0}\%    \\
                & Other           & \num{2.0}\%     & \num{3.0}\%     \\
\midrule
Donor Diabetes History & No       & \num{79.0}\%    & \num{73.0}\%    \\
                       & Yes, 0-5 years & \num{10.0}\% & \num{10.0}\% \\
                       & Other    & \num{12.0}\%    & \num{17.0}\%    \\
\midrule
Insulin Dependent Donor & No      & \num{62.0}\%    & \num{52.0}\%    \\
                       & Unknown  & \num{9.0}\%     & \num{12.0}\%    \\
                       & Yes      & \num{30.0}\%    & \num{36.0}\%    \\
\midrule
Donor Hypertension History & Yes  & \num{67.0}\%    & \num{70.0}\%    \\
                           & No    & \num{31.0}\%    & \num{28.0}\%    \\
                           & Other & \num{2.0}\%     & \num{2.0}\%     \\
\midrule
Donor Cancer History & No          & \num{94.0}\%    & \num{92.0}\%    \\
                     & Yes         & \num{5.0}\%     & \num{7.0}\%     \\
                     & Other       & \num{1.0}\%     & \num{2.0}\%     \\
\midrule
Donor CMV     & Positive    & \num{66.0}\%    & \num{62.0}\%    \\
                     & Negative    & \num{33.0}\%    & \num{37.0}\%    \\
\midrule
Donor HepB Surface Antigen & Negative & \num{100.0}\% & \num{100.0}\% \\     & Positive & \num{0.0}\%   & \num{0.0}\%   \\
\midrule
Donor HepB Core Antibody   & Negative & \num{92.0}\%  & \num{94.0}\%  \\
                                   & Positive & \num{8.0}\%   & \num{6.0}\%   \\
\midrule
Donor HepC Antibody        & Negative & \num{66.0}\%  & \num{95.0}\%  \\
                                   & Positive & \num{31.0}\%  & \num{5.0}\%   \\
                                   & Other    & \num{3.0}\%   & \num{0.0}\%   \\
\midrule
Donor Tattoos Presence            & No       & \num{69.0}\%  & \num{69.0}\%  \\
                                   & Yes      & \num{31.0}\%  & \num{31.0}\%  \\
\midrule 
Donor DCD PBD  & No       & \num{91.0}\%  & \num{91.0}\%  \\
                                   & Yes      & \num{9.0}\%   & \num{9.0}\%   \\         
\bottomrule
\end{tabular}
}
\end{minipage}%
\hspace{0.01\textwidth}
\begin{minipage}{0.49\textwidth}
\resizebox{\linewidth}{!}{%
\begin{tabular}{llSS}
\toprule
                 &                           & \multicolumn{1}{c}{Accept} & \multicolumn{1}{c}{Reject} \\ \toprule
Donor Smoking History             & No       & \num{69.0}\%  & \num{60.0}\%  \\
                                   & Yes      & \num{31.0}\%  & \num{37.0}\%  \\
                                   & Other    & \num{0.0}\%   & \num{3.0}\%   \\
\midrule
Donor MI History                  & No       & \num{92.0}\%  & \num{88.0}\%  \\
                                   & Yes      & \num{6.0}\%   & \num{10.0}\%  \\
                                   & Other    & \num{2.0}\%   & \num{3.0}\%   \\
\midrule
Donor Cocaine Usage History       & No       & \num{79.0}\%  & \num{79.0}\%  \\
                                   & Yes      & \num{20.0}\%  & \num{18.0}\%  \\
                                   & Other    & \num{2.0}\%   & \num{2.0}\%   \\
\midrule
Donor IV Drug Usage History       & No       & \num{92.0}\%  & \num{93.0}\%  \\
                                   & Yes      & \num{7.0}\%   & \num{5.0}\%   \\
                                   & Other    & \num{1.0}\%   & \num{2.0}\%   \\
\midrule
Donor Other Drug Usage History    & No       & \num{64.0}\%  & \num{63.0}\%  \\
                                   & Yes      & \num{35.0}\%  & \num{35.0}\%  \\
                                   & Other    & \num{1.0}\%   & \num{2.0}\%   \\
\midrule
Donor Insulin Use                 & No       & \num{54.0}\%  & \num{55.0}\%  \\
                                   & Yes      & \num{46.0}\%  & \num{45.0}\%  \\
\midrule
Donor CDC Risk HIV                & No       & \num{87.0}\%  & \num{87.0}\%  \\
                                   & Yes      & \num{13.0}\%  & \num{13.0}\%  \\
\midrule
Donor Urine Protein               & Yes      & \num{51.0}\%  & \num{57.0}\%  \\
                                   & No       & \num{49.0}\%  & \num{42.0}\%  \\
                                   & Other    & \num{0.0}\%   & \num{1.0}\%   \\
\midrule
Donor HCV NAT              & Negative & \num{95.0}\%  & \num{97.0}\%  \\
                                   & Positive & \num{5.0}\%   & \num{3.0}\%   \\
\midrule
Donor Antihypertensive Use        & No       & \num{70.0}\%  & \num{72.0}\%  \\
                                   & Yes      & \num{30.0}\%  & \num{28.0}\%  \\
\midrule
Donor Arginine Use                & No       & \num{55.0}\%  & \num{55.0}\%  \\
                                   & Yes      & \num{45.0}\%  & \num{45.0}\%  \\
\midrule
Donor Coronary Angiography        & No       & \num{78.0}\%  & \num{85.0}\%  \\
                                   & Yes      & \num{22.0}\%  & \num{15.0}\%  \\
\midrule
Donor Legally Brain Dead          & Yes      & \num{73.0}\%  & \num{66.0}\%  \\
                                   & No       & \num{27.0}\%  & \num{34.0}\%  \\
\midrule
Donor Interstitial Fibrosis       & Occasional & \num{40.0}\% & \num{32.0}\% \\
                                   & Some & \num{28.0}\% & \num{27.0}\% \\
                                   & Most & \num{27.0}\% & \num{24.0}\% \\
                                   & Other  & \num{5.0}\%   & \num{17.0}\%   \\
\midrule
Time of Day                       & Late Night & \num{19.0}\%  & \num{23.0}\%  \\
                                   & Noon      & \num{19.0}\%  & \num{16.0}\%  \\
                                   & Eve       & \num{18.0}\%  & \num{16.0}\%  \\
                                   & Early Morning & \num{18.0}\%  & \num{16.0}\%  \\
                                   & Morning   & \num{15.0}\%  & \num{16.0}\%  \\
                                   & Night     & \num{11.0}\%  & \num{14.0}\%  \\
\bottomrule
\end{tabular}
}
\end{minipage}%

\end{table}

\sisetup{round-mode=places, round-precision=1, table-format=2.2, table-number-alignment=center, round-pad=false}

\begin{table}[htbp]
\caption{Deceased donor kidney offers between 2016 and 2021 stratified by offer response: Continuous features.}
\label{tbl:basic_stat_recipient_cont}
\centering

\begin{minipage}{0.8\textwidth}
\resizebox{\linewidth}{!}{
\begin{tabular}{llSS}
\toprule
                 &                           & \multicolumn{1}{c}{Accept} & \multicolumn{1}{c}{Reject} \\ \toprule
Distance from Donor to Center (miles)        &                           & \num{623.41} \text{ (883.24)} & \num{1340.22} \text{ (1182.68)} \\[1pt]
Cold Ischemia Time (minutes)                 &                           & \num{348.36} \text{ (2856.18)} & \num{575.23} \text{ (603.27)} \\[1pt]
Kidney Donor Risk Index (RAO)                &                           & \num{1.79} \text{ (0.10)} & \num{1.82} \text{ (0.10)} \\[1pt]
Donor Age (years)                            &                           & \num{55.26} \text{ (8.45)} & \num{55.57} \text{ (8.36)} \\[1pt]
Donor Height (cm)                            &                           & \num{168.15} \text{ (11.92)} & \num{169.15} \text{ (12.97)} \\[1pt]
Donor Weight (kg)                            &                           & \num{84.34} \text{ (23.26)} & \num{88.56} \text{ (25.21)} \\[1pt]
Donor Creatinine Level (mg/dL)               &                           & \num{1.35} \text{ (1.14)} & \num{1.79} \text{ (1.44)} \\[1pt]
Nominal GDP of Center State (USD)    &                           & \num{80494.52} \text{ (25626.32)} & \num{81012.87} \text{ (24054.53)} \\[1pt]
Donor BMI (kg/m$^2$)                           &                           & \num{29.68} \text{ (7.34)} & \num{30.75} \text{ (7.80)} \\[1pt]
Donor Blood Urea Nitrogen (mg/dL)            &                           & \num{24.77} \text{ (18.07)} & \num{30.95} \text{ (20.74)} \\[1pt]
Donor Death Mechanism                        &                           & \num{24.44} \text{ (123.64)} & \num{22.31} \text{ (116.43)} \\[1pt]
Kidney Glomeruli Count                  &                           & \num{59.08} \text{ (36.62)} & \num{59.87} \text{ (37.19)} \\[1pt]
1-Year Higher CIT Acceptances                &                           & \num{12.35} \text{ (18.83)} & \num{5.74} \text{ (11.95)} \\[1pt]
2-Year Higher CIT Acceptances                &                           & \num{21.32} \text{ (31.10)} & \num{10.43} \text{ (21.35)} \\[1pt]
3-Year Higher CIT Acceptances                &                           & \num{27.86} \text{ (38.74)} & \num{13.96} \text{ (28.02)} \\[1pt]
1-Year Higher KDRI Acceptances               &                           & \num{5.95} \text{ (9.42)} & \num{3.15} \text{ (5.49)} \\[1pt]
2-Year High KDRI Acceptances                 &                           & \num{10.15} \text{ (15.80)} & \num{5.73} \text{ (9.77)} \\[1pt]
3-Year High KDRI Acceptances                 &                           & \num{12.90} \text{ (18.84)} & \num{7.52} \text{ (12.53)} \\[1pt]
Center Distance to Medium Airport (miles)    &                           & \num{21.97} \text{ (78.28)} & \num{19.23} \text{ (49.51)} \\[1pt]
Center Distance to Large Airport (miles)     &                           & \num{61.78} \text{ (113.96)} & \num{59.57} \text{ (100.59)} \\[1pt]
Center Patient Count                         &                           & \num{1372.96} \text{ (896.46)} & \num{1194.74} \text{ (844.37)} \\[1pt]
Donor Distance to Medium Airport (miles)     &                           & \num{31.55} \text{ (152.54)} & \num{28.62} \text{ (130.81)} \\[1pt]
Donor Distance to Large Airport (miles)      &                           & \num{98.47} \text{ (178.99)} & \num{90.67} \text{ (158.90)} \\[1pt]
Center Acceptance Rate                       &                           & \num{0.04} \text{ (0.02)} & \num{0.03} \text{ (0.02)} \\[1pt]
Center Average Accepted KDRI                 &                           & \num{1.24} \text{ (0.10)} & \num{1.20} \text{ (0.10)} \\[1pt]
Center Average Accepted Age (years)          &                           & \num{49.01} \text{ (2.15)} & \num{49.09} \text{ (3.33)} \\[1pt]
2-Year Older Kidney Acceptances              &                           & \num{9.38} \text{ (21.93)} & \num{5.69} \text{ (14.24)} \\[1pt]
2-Year Higher Creatinine Acceptances         &                           & \num{32.70} \text{ (41.56)} & \num{19.80} \text{ (31.80)} \\[1pt]
2-Year Diabetes-Related Acceptances          &                           & \num{78.52} \text{ (73.26)} & \num{57.59} \text{ (60.24)} \\[1pt]
2-Year Drug-Related Acceptances              &                           & \num{93.19} \text{ (72.04)} & \num{73.70} \text{ (59.97)} \\[1pt]
2-Year DCD Acceptances                       &                           & \num{55.42} \text{ (59.07)} & \num{39.35} \text{ (46.18)} \\[1pt]
2-Year KAP Eligible Acceptances              &                           & \num{0.05} \text{ (0.40)} & \num{0.01} \text{ (0.28)} \\[1pt]
\bottomrule
\end{tabular}}

\end{minipage}%

\end{table}

A significant challenge during data processing was the imbalanced nature of the dataset, with kidney rejections vastly outnumbering acceptances at a ratio of approximately 34 to 1. This imbalance poses difficulties in accurately modeling and predicting acceptance outcomes, and it can lead to biased model predictions and limit the insights of our ranking model. To address this, we employ the following data censoring methods: 
\begin{itemize}\setItemSep{0.3em}
    \item For accepted kidneys, we exclude offers to centers that has a lower ratio of offers to total patient count than the accepting center. This approach ensures we focus on centers that are comparable in their offer response to the accepting center. For example, if a center accepts a kidney after receiving offers for 20\% of its patient population, then any other center that rejects the same kidney without receiving offers for at least 20\% of its population is excluded from our analysis (for that particular kidney). The rationale is that these latter centers might have accepted the kidney if they had received offers for a comparable proportion of their patient population.
    
    \item For unused kidneys, we retain a representative subset by censoring offers using a threshold based on the ratio of offers to the total patient count. Our target was to ensure that acceptances constituted approximately 5\% of the total dataset.
\end{itemize}

This approach helps in creating a dataset that aligns with real-world scenarios while being conducive to ML analysis. Table~\ref{tab:dataset_summary} provides summary statistics on the censored datasets according to different data specifications.

%%%%%%%%%%%%%%%%%%%%%%%%%%%%%%%%%%%%%%%%%%%%%%%
\subsubsection{Data Exclusion}\label{sc:data_exc}
%%%%%%%%%%%%%%%%%%%%%%%%%%%%%%%%%%%%%%%%%%%%%%%
To ensure the quality and reliability of the dataset, we excluded data points missing critical values. First, we removed donors from the deceased donor dataset that were missing clamp dates, resulting in the exclusion of approximately 0.005\% of unique donors. Next, we eliminated any kidney offers in our dataset that did not have a corresponding entry in the deceased donor dataset, which accounted for approximately 12.18\% of the unique donors referenced in the PTR dataset. Additionally, we excluded cases where a kidney was accepted more than twice, which represented about 5\% of the remaining unique donors. After the exclusions and data losses, the dataset contained 61,794 unique donors.

% The usage of numerous features in our analysis is not a concern due to the application of the LightGBM model. 
% This model is well-suited for handling high-dimensional data since as a tree model it effectively performs feature selection.

\setlength{\tabcolsep}{1.5pt}
\renewcommand{\arraystretch}{1.25}
\begin{table}[!ht]
\centering
\caption{Summary of the dataset variants used in the empirical analysis}
\label{tab:dataset_summary}
% \begin{tabular}{lllll}
\resizebox{0.9\textwidth}{!}{
\begin{tabular}{P{0.22\textwidth} C{0.12\textwidth} C{0.18\textwidth} C{0.20\textwidth} C{0.20\textwidth}}
\toprule 
& \textbf{\# of centers} & \textbf{\# of donors} & \textbf{\# of rejected offers} &  \textbf{\# of accepted offers}\\
% \textbf{}                                           & \textbf{\# of centers} & \textbf{\# of donors} & \textbf{\begin{tabular}[c]{@{}l@{}}\# of rejected \\ offers\end{tabular}} & \textbf{\begin{tabular}[c]{@{}l@{}}\# of accepted \\ offers\end{tabular}} \\ 
\midrule
Uncensored dataset & 250 & 61,794 & 1,701,656 & 50,386 \\ 
Censored dataset & 250 & 52,470 & 1,005,725 & 50,386 \\ 
1.65 $\le$ KDRI $\le$ 2.0 & 247 & 7,954 & 1,748,025 & 4,017 \\ 
1.65 $\le$ KDRI $\le$ 2.0 and censored & 215 & 5,591 & 77,609 & 3,944 \\ 
\bottomrule
\end{tabular}
}
\end{table}

%%%%%%%%%%%%%%%%%%%%%%%%%%%%%%%%%%%%%%%%%%%%%%%%%%%%%%%%%%%%%%%%%%%%%%
% \subsection{Baseline Methods for Kidney Allocation}
\subsection{Intervention Methods for Allocating Hard-to-Place Kidneys}
\label{blm}
%%%%%%%%%%%%%%%%%%%%%%%%%%%%%%%%%%%%%%%%%%%%%%%%%%%%%%%%%%%%%%%%%%%%%%

We develop and test several data-driven heuristic approaches to create an effective intervention framework for allocating hard-to-place kidneys to centers with a higher likelihood of accepting them. The primary goal of these interventions is to allocate hard-to-place kidneys in a timely manner, preventing prolonged CIT during the offer process, which could lead to organs being unused. To evaluate the performance of different intervention approaches, we calculate the number of offer rejections (non-accepting centers) encountered before reaching an accepting center, which we refer to as the \textit{number of centers seen (NCS)} score. This metric is straightforward and easy to interpret, providing a solid basis for comparing the effectiveness of interventions. As summarized in \Cref{sc:Intro}, in the current organ allocation system, each final offer sent to a transplant center can take up to an hour, potentially contributing to higher CIT. Therefore, a low NCS score indicates a timely placement of hard-to-place kidneys, minimizing CIT and reducing the risk of non-use, while a high score suggests longer CIT and a greater risk of the organ being unused.

Given a set of $n$ kidneys $\{K_1, K_2, \hdots, K_n\}$, with each kidney $K_i$ offered to $c_i$ centers before acceptance, we define the NCS as the average number of offers over $n$ kidneys as follows:
    \begin{equation}
    \label{eq:baseline}
    \text{NCS} = \frac{1}{n}\sum_{i=1}^{n}c_i.
    \end{equation}

Next, we summarize the three heuristic approaches we propose for allocating hard-to-place kidneys, along with a baseline representing current organ allocation practices.

\begin{itemize}\setItemSep{0.3em}
\item \textit{Baseline}: This value represents the ground truth, corresponding to the actual order of offers as they previously occurred in the real-world data. It reflects the current practice in organ allocation and will be used as the benchmark for evaluating the effectiveness of other intervention methods. By comparing our proposed heuristic approaches to this baseline, we can assess their potential improvements over existing practices in terms of reducing CIT and improving timely kidney placement.

\item \textit{Acceptance Likelihood Prediction (ALP) Framework}: This data-driven framework employs an ML model trained on our constructed dataset to predict the likelihood of each center accepting a given donor kidney. These predicted probabilities are then used to rank the centers, generating an ordered list for each kidney, denoted as $\mathcal{O} = \{o_1, o_2, \ldots, o_m\}$, where $o_j$ represents the $j$-th center in the offering sequence. This ranking order aims to prioritize centers with a higher predicted likelihood of acceptance, thereby potentially reducing the number of offers needed before a kidney is accepted.

    % Do we still want to include this type of removal metric??
    % \item \textit{Heuristic approach 1} ($H_1$): This heuristic maintains the current order of centers $O$ but excludes centers that have not accepted a kidney with a higher Kidney Donor Risk Index (KDRI) in the past two years. Let $A$ be the set of such centers. Then the order suggested by this heuristic is given by:
    % \begin{equation}
    % H_1 = O \setminus A.
    % \end{equation}

    \item \textit{KDRI-based Heuristic}: KDRI is widely used as a proxy for organ quality in both clinical practice and academic literature. Centers with a history of accepting kidneys with a high KDRI are likely better equipped, either through specialized expertise or advanced infrastructure, to manage such transplants~\citep{rao2009comprehensive}. Additionally, these centers may serve patient populations that prioritize immediate transplantation over waiting for potentially lower-risk organs, even if the KDRI is slightly elevated. We propose a heuristic approach that orders transplant centers based on the number of accepted kidneys that are higher KDRI than the current offer over the past two years. %\mycomment{S: We need to clarify what we mean by past acceptance, there are various ways to define this. Is this just the count?} \textcolor{blue}{changed the text to provide more clarity}. 
    This method aims to prioritize centers that have demonstrated a willingness and capability to handle hard-to-place kidneys of varying quality. The ordered list of centers, denoted as $\mathcal{P} = \{p_1, p_2, \hdots, p_m\}$, reflects the likelihood of acceptance based on historical behavior.
    
    \item \textit{KAP-based Heuristic}: The KAP program aims to make more efficient use of kidneys that might otherwise go unused~\citep{mohan2022accelerating}. This heuristic approach ranks transplant centers based on the number of kidneys they have accepted in the past two years that qualify them for KAP (see \Cref{tab:kap} for details). The rationale behind this approach is that centers with higher KAP-eligible acceptances are likely more adept at managing and utilizing such kidneys efficiently. We denote the sequence of centers ranked by their number of KAP-eligible acceptances as $\mathcal{Q} = \{q_1, q_2, \hdots, q_m\}$. By utilizing this ranking, the heuristic seeks to prioritize offers to centers with a demonstrated willingness to accept kidneys that are conventionally hard to place but are still viable for transplantation. 
\end{itemize}

\setlength{\tabcolsep}{6pt}
\renewcommand{\arraystretch}{1.25}
\begin{table}[b]
\centering
\caption{An illustration of the three proposed heuristics over a sample kidney that was offered to 5 different centers before being accepted}
\label{tab:example}
\resizebox{0.85\textwidth}{!}{
\begin{tabular}{lrcccc}
% {P{0.005\textwidth}C{0.15\textwidth}C{0.148\textwidth}C{0.143\textwidth}C{0.153\textwidth}C{0.171\textwidth}}
% \begin{tabularx}{\textwidth}{|X|X|c|X|X|X|X|}
\toprule
\multirow{2}{*}{Index} & \multirow{2}{*}{Center} & \multirow{2}{*}{Offer Accept} & \multirow{2}{*}{KDRI\_2\_year} & \multirow{2}{*}{KAP\_eligible} & Acceptance \\[-2pt]
&&&&& Probability\\
\toprule
1 & 23033  & 0 & 23 & 1 & 0.037 \\
2 & 15438  & 0 & 104 & 2 & 0.208 \\
3 & 22568  & 0 & 13 & 1 & 0.018 \\
4 & 682  & 0 & 44 & 2 & 0.118 \\
5 & 3937  & 1 & 25 & 0 & 0.580 \\
\bottomrule
% \end{tabularx}
\end{tabular}
}
\end{table}
% This is likely best understood with an example. 
We illustrate the aforementioned intervention mechanisms with an example. Table~\ref{tab:example} demonstrates how each proposed framework evaluates a sample kidney that was offered to five different centers before being accepted. The ``index" column represents the actual sequence of offers as observed in the data, serving as the ground truth for comparison. The table includes several key metrics related to the kidney offers to different centers. ``Center Code" refers to the unique anonymized identifier for each center. ``Offer Accept" is a binary indicator showing whether the offer was accepted by the center. ``KDRI\_2\_year" indicates the number of higher KDRI kidneys that a center has accepted in the past two years. ``KAP\_eligible" shows the number of kidneys accepted by the center that qualify it for the KAP program. Lastly, ``Acceptance Probability" provides the ML model-predicted likelihood that the center will accept the kidney. 

Table~\ref{tab:examplecont} presents the ranked orderings of each center according to the proposed mechanisms. Under the KDRI-based heuristic, centers are ordered based on the number of high KDRI kidneys they have accepted in the past two years. The KAP heuristic ranks centers by the number of kidneys they have accepted that qualify for accelerated placement under the KAP program. Finally, the ALP framework orders centers by their predicted acceptance probabilities. The ground truth shows that four centers rejected the organ before it was accepted, resulting in an NCS score of 4. The ranking based on the KDRI-based heuristic shows that two centers would reject the organ before acceptance, yielding an NCS score of 2. Similarly, the KAP-based heuristic also results in an NCS score of 4. In contrast, the ALP framework achieves the lowest NCS score of 0, indicating that the first center in the predicted order would accept the kidney.

% Note that in the event of a tie in ordering the ordering defaults to the original order. 

\setlength{\tabcolsep}{1.5pt}
\renewcommand{\arraystretch}{1.25}
\begin{table}[t]
\centering
\caption{Order suggested via the different methods.}
% \begin{tabular}{|c|c|c|c|c|}
\resizebox{0.65\textwidth}{!}{
\begin{tabular}{C{0.15\textwidth}C{0.148\textwidth}C{0.143\textwidth}C{0.153\textwidth}}
\toprule
Order Baseline & Order KDRI Heuristic & Order KAP Heuristic & Order ALP Framework  \\
\midrule
1 & 2 & 2 & \textbf{5}  \\
2 & 4 & 4 & 2  \\
3 & \textbf{5} & 1 & 4  \\
4 & 1 & 3 & 1  \\
\textbf{5} & 3 & \textbf{5} & 3  \\
\bottomrule
\end{tabular}
}
\label{tab:examplecont}
\end{table}

% Figure~\ref{fig:flow} provides a flow diagram for the ranking metric calculation.
% \begin{figure}[!ht]
%     \centering
%     \includegraphics[scale=0.5]{figures/organ.png}
%     \caption{Flow diagram of the ranking metric.}
%     \label{fig:flow}
% \end{figure}

%%%%%%%%%%%%%%%%%%%%%%%%%%%%%%%%%%%%%%%%%%%%%%%%%%%%%%%%%%%%%%%%%%%%%%%%%%%%%%%%%

%%%%%%%%%%%%%%%%%%%%%%%%%%%%%%%%%%%%%%%%%%%%%%%%%%%%%%%%%%%%%%%%%%%%%%
\subsection{ML Models Used in the ALP Framework}\label{sc:ML_models}
%%%%%%%%%%%%%%%%%%%%%%%%%%%%%%%%%%%%%%%%%%%%%%%%%%%%%%%%%%%%%%%%%%%%%% 

We utilize ML models for predicting the probability of acceptance by a center for a given kidney. %, (2) ranking centers based on their likelihood of accepting an offered kidney.
% For the first task, 
Specifically, we train ML models by splitting the dataset donor-wise into training and testing sets and then train the models to predict acceptance events. 
% For the second task, we 
We then use the raw probability predictions from these trained models to rank the centers for each donor, prioritizing those with the highest likelihood of acceptance. To evaluate the ranking performance of the ML models, we employ the method described in Section~\ref{blm}.

% This allows us to determine if the model or indeed a heuristic gives a better ranking than the ground truth. In what follows, we briefly discuss the ML models used in our analysis. 

Given the imperative for both high accuracy and interpretability in predicting organ acceptance, tree-based models are the prevalent choice in the literature \citep{ashiku2021machine,guijo2021statistical}. Building on this, we employ a range of ML models based on prior studies, while also incorporating more recent gradient boosting models to leverage their superior predictive capabilities and efficiency in handling large and complex datasets.
% We use \textit{Logistic Regression} as a baseline ML method, as it models binary outcomes through linear relationships. A \textit{Decision Tree} is included for its simplicity and ease of interpretation. To enhance predictive performance, 
Specifically, we train a \textit{Random Forest} model, an ensemble of decision trees that provides improved accuracy~\citep{pedregosa2011scikit}. The \textit{Extra Trees} model, similar to Random Forest, uses multiple decision trees but incorporates more randomized feature splits, reducing the risk of overfitting~\citep{goetz2014extremely}. Additionally, we include various gradient boosting models. \textit{Light Gradient Boosting Machine (LGBM)} is chosen for its effectiveness in managing high-dimensional feature spaces and large datasets~\citep{ke2017lightgbm}. \textit{CatBoost} is selected for its capability to handle categorical data efficiently without extensive preprocessing~\citep{dorogush2018catboost}. 
Additionally, we use \textit{Logistic Regression} as a baseline ML method, as it models binary outcomes through linear relationships. A \textit{Decision Tree} model is also included for its simplicity and ease of interpretation.

Together, these models represent a broad spectrum of ML techniques, each with unique strengths and  applications. Their performances, as quantified in our study, can guide practitioners in selecting the most appropriate model for their specific kidney allocation intervention tasks.

\subsection{ML Interpretability Methods}
%%%%%%%%%%%%%%%%%%%%%%%%%%%%%%%%%%%%%%%%%%%%%%%%%%%%%%%%%%%%%%%%%%%%%% 
The black-box nature of many ML models necessitates actionable and interpretable insights into their decisions. Ensuring transparency is crucial not only for gaining acceptance of these models but also for carefully aiding decisions in domains like organ allocation, where actions have life-altering implications. In our analysis, we use SHAP (SHapley Additive exPlanations) and its variants, particularly, \textit{TreeSHAP}, to elucidate model predictions and provide clear, interpretable explanations~\citep{lundberg2018consistent}.

TreeSHAP is an interpretability method specifically designed for explaining the outputs of tree-based ML models. This method bridges game theory and local explanations by calculating a feature's Shapley value, which represents the ``fair'' contribution of that feature to the model's output. This Shapley value is determined by averaging the marginal contributions across all possible feature permutations. TreeSHAP leverages the inherent structure of tree models to compute these game-theoretic values in a highly efficient manner, making it well-suited for practical applications.

The SHAP library offers methods for both local and global interpretability for ML models. Locally, it provides insight into how individual features or combinations of features influence a single prediction. In the context of the kidney allocation interventions problem, local explanations can elucidate the model's estimated acceptance probability for a specific donor-center pair. Globally, the aggregate of SHAP values helps us understand the overal behavior of the model, identifying which features are most influential across all predictions. This dual-level insight is invaluable, not only for uncovering previously misunderstood factors affecting kidney allocation interventions but also for providing confidence in model decisions. These insights are essential for validating and fine-tuning the applications of ML models in high-stakes situations like organ allocation.

%%%%%%%%%%%%%%%%%%%%%%%%%%%%%%%%%%%%%%%%%%%%%%%%%%%%%%%%%%%%%%%%%%%%%%%%%%%%%%%%%
\section{Results}
\label{res}
In this section, we present the findings from our comprehensive numerical study using various ML models and heuristics for the kidney allocation problem. Our analysis covers different compositions of kidney allocation data to showcase the capabilities of these models for this task. We begin by evaluating the performance of different ML models on a subset of the dataset for model selection, focusing on their effectiveness in the kidney allocation interventions.
We then extend the analysis to the uncensored (i.e., full) dataset, assessing the models in greater detail using standard classification metrics as well as the NCS score. Following this, we discuss the results obtained from the censored dataset, including an explanation of the censoring process, performance different ranking approaches under these conditions, and the associated performance metrics. Finally, we present results from filtered datasets based on different ranges of the KDRI, accompanied by local and global explanations of the model predictions to provide deeper insights into the factors influencing these decisions.

%%%%%%%%%%%%%%%%%%%%%%%%%%%%%%%%%%
\subsection{ML Model Selection}
%%%%%%%%%%%%%%%%%%%%%%%%%%%%%%%%%%
To identify the best performing ML model, we conduct an analysis using ten random splits from the data for training and testing. We compare the performance of several ML models based on the Macro F1 score and the average number of offers before acceptance, as indicated by the NCS score.
The considered ML models include CatBoost, Decision Tree, LGBM, Logistic Regression, Random Forest, 
% \mycomment{S: Include kNN in \Cref{sc:ML_models}.} \textcolor{blue}{We do not reference them anywhere else} \mycomment{S: Table 6 reports kNN results, but we can delete that.}\textcolor{blue}{Removed.}, 
and Extra Trees. % (see \Cref{sc:ML_models} for details). 
% These models represent a diverse range of methodologies, including distance-based algorithms, tree-based methods, and ensemble techniques. 
The results of this comparison are presented in Table~\ref{tab:model_comparison}.

\setlength{\tabcolsep}{3.0pt}
\renewcommand{\arraystretch}{1.15}
\begin{table}[!ht]
\centering
\caption{Performance comparison of various ML models within the ALP framework. Standard deviations are presented in parentheses. The best-performing model(s) for each performance metric is highlighted in bold.}
\label{tab:model_comparison}
\resizebox{0.709\textwidth}{!}{
\begin{tabular}{lcccccc}
\toprule
Model &         NCS &        Macro F1 &    Accuracy &       Recall &    Precision \\
\midrule
\multirow{2}{*}{CatBoost} & 0.077 & 0.725 & \textbf{0.969} & 0.382 & 0.607 \\
 & (0.028) & (0.036) & (0.007) & (0.053) & (0.120) \\ 
 \midrule
\multirow{2}{*}{Decision Tree} & 0.353 & 0.592 & 0.906 & 0.405 & 0.167 \\
 & (0.135) & (0.024) & (0.016) & (0.054) & (0.039) \\ 
 \midrule
% \multirow{2}{*}{K-Nearest Neighbors} & 0.313 & 0.595 & 0.966 & 0.126 & 0.588 \\
%  & (0.146) & (0.021) & (0.004) & (0.026) & (0.121) \\
%  \midrule
\multirow{2}{*}{LGBM} & \textbf{0.055} & \textbf{0.803} & 0.968 & \textbf{0.550} & 0.722 \\
 & (0.015) & (0.010) & (0.002) & (0.021) & (0.037) \\
\midrule
\multirow{2}{*}{Logistic Regression} & 0.687 & 0.511 & 0.965 & 0.021 & 0.702 \\
 & (0.121) & (0.012) & (0.004) & (0.013) & (0.277) \\
 \midrule
\multirow{2}{*}{Random Forest} & 0.111 & 0.626 & \textbf{0.969} & 0.160 & \textbf{0.837} \\
 & (0.046) & (0.028) & (0.004) & (0.035) & (0.145) \\
 \midrule
\multirow{2}{*}{Extra Trees} & 0.145 & 0.612 & 0.968 & 0.146 & 0.714 \\
 & (0.029) & (0.026) & (0.004) & (0.034) & (0.121) \\
\bottomrule
\end{tabular}
}
\end{table}

% \multirow{2}{*}{Support Vector Machine} & 0.770 & 0.491 & 0.965 & 0.000 & 0.000 \\
%  & (0.129) & (0.001) & (0.004) & (0.000) & (0.000) \\\hline

\begin{figure}[b]
    \centering
    \includegraphics[width = .8\textwidth]{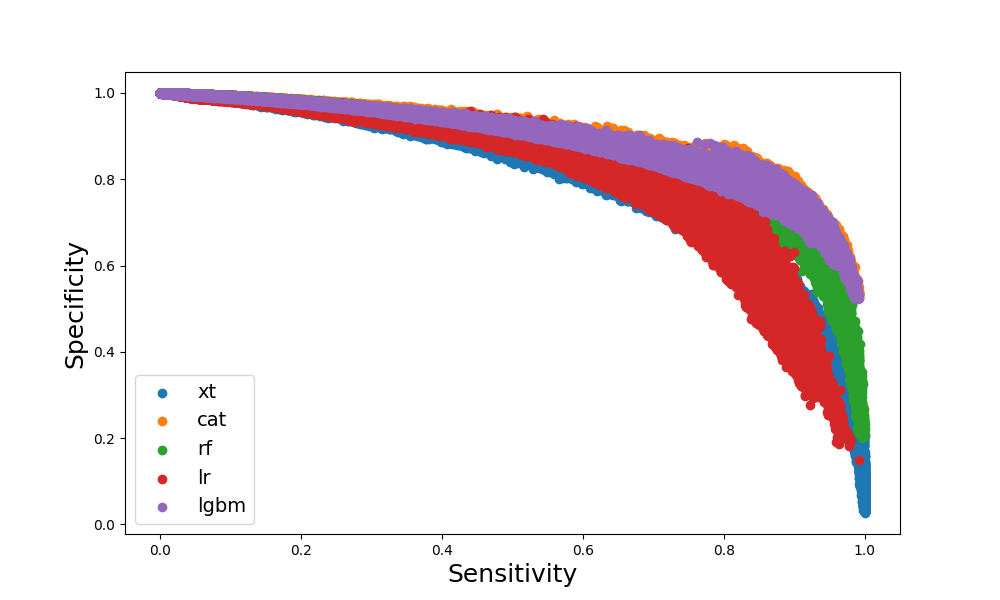}
    \caption{Comparison of models' sensitivity-specificity plots %\highlight{\small 1pt smaller. Take this as a reference. } \highlight{\footnotesize 2pt smaller. This is still okay, but anything smaller is too small. Currently, axis labels are smaller than this.}\textcolor{blue}{Noted} 
    } 
    \label{fig:ss_full}
\end{figure}

Table~\ref{tab:model_comparison} shows that the LGBM model outperforms others in terms of both Macro F1 score and the NCS score. To further elucidate the performance of the selected models, we compare the sensitivity-specificity plots across varying decision thresholds. These curves are generated by adjusting the models at different cut-off data censoring ratios. As shown in Figure~\ref{fig:ss_full}, both the LGBM and CatBoost models performs well, with CatBoost occasionally surpassing LGBM in regions of higher sensitivity.

% \mycomment{S: Move to appendix}
% \textcolor{blue}{Moved to appendix}

% \begin{figure}[!ht]
%     \centering
%     \includegraphics[width = .8\textwidth]{figures/sens_specs/ss_sample.png}
%     \caption{Comparison of Models Sensitivity Specificity Plots (Sample)}
%     \label{fig:ss_sample}
% \end{figure}

\begin{figure}[!ht]
    \centering
    \includegraphics[width = .8\textwidth]{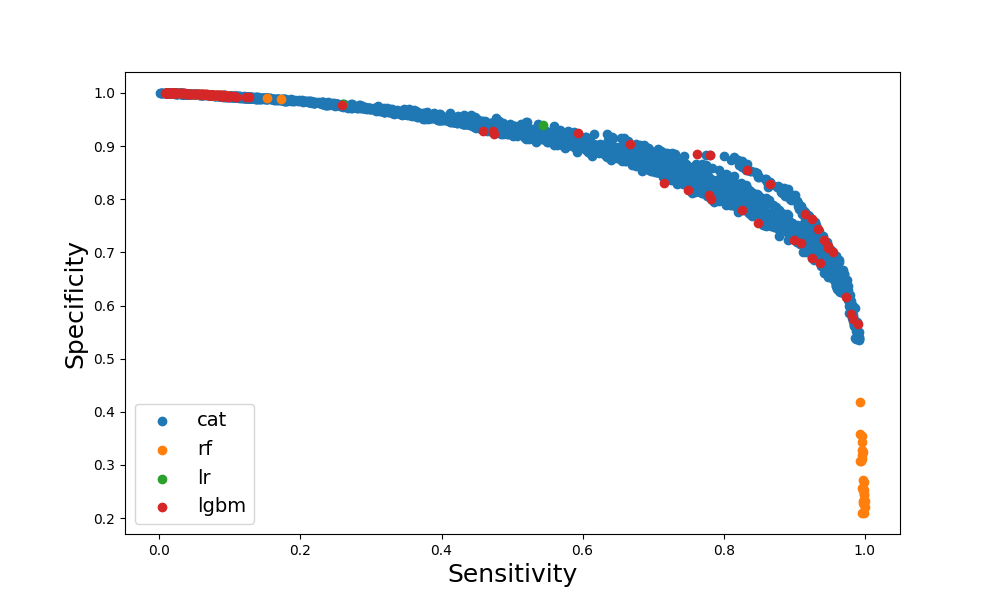}
    \caption{Comparison of models' sensitivity specificity plots showing non-dominated results 
    % \mycomment{S: Can we report the percent of the dots belonging to each model. Because of the overlaps, it is not clear which models dominate the graph.}\textcolor{blue}{Added at paragraph referencing this figure}
    }
    \label{fig:ss_dom}
\end{figure}

\Cref{fig:ss_dom} provides a more nuanced view by highlighting models that maintain competitive (non-dominated) performance across the entire sensitivity-specificity spectrum. 
% In our analysis, the non-dominated proportion of each model type within the ALP framework was calculated as follows: xt had a proportion of 0.0\%, cat had a proportion of 93.90\%, rf had a proportion of 1.69\%, lr had a proportion of 0.07\%, and lgbm had a proportion of 4.34\%. 
While the LGBM model consistently shows strong performance, it is important to note that CatBoost exhibits comparable, and in some instances, superior performance in certain regions of the spectrum. 

Both LGBM and CatBoost offer robust frameworks for in-depth model interpretation, enabling a clear understanding of how individual features contribute to predictions. This interpretability is particularly critical in high-stakes domains like organ allocation, where transparency in the decision-making process is essential for informed and ethical choices. While both models demonstrate strong performance, we ultimately select the LGBM model as our preferred approach for the kidney allocation intervention problem due to its slight advantage in predictive accuracy and interpretability. Specifically, LGBM excels in ranking centers, delivers consistent performance across various metrics, and offers clear, interpretable insights; making it the optimal choice for this task.

%%%%%%%%%%%%%%%%%%%%%%%%%%%%%%%%%%%%%%%%%%%%%%%%%%%%%%%
\subsection{Performance of the ALP Framework over All Kidneys}\label{sc:ALP_performance_unrestricted}
%%%%%%%%%%%%%%%%%%%%%%%%%%%%%%%%%%%%%%%%%%%%%%%%%%%%%%%
We evaluate the performance of the ALP framework using our comprehensive kidney allocation dataset, including all kidneys without restricting to hard-to-place cases. The objective of this analysis is to assess how effectively ALP identifies transplant centers with the highest likelihood of accepting any given kidney. We perform an 80/20 train-test split on the donors data, train an LGBM model on the training set, and then report the model's performance on the test set.

\begin{table}[b]
\centering
\caption{Classification report for LGBM - uncensored data}
\label{tab:full_classification_report}
\begin{tabular}{lrrrr}
\toprule
  & Precision & Recall & F1 score & Support \\
\midrule
0 (Reject) & 0.972 & 0.999 & 0.985 & 342,383 \\
1 (Accept) & 0.220 & 0.007 & 0.013 & 10,070 \\
\midrule
Accuracy & 0.971 & 0.971 & 0.971 & 352,453 \\
Macro avg & 0.596 & 0.503 & 0.499 & 352,453 \\
Weighted avg & 0.950 & 0.971 & 0.957 & 352,453 \\
\bottomrule
\end{tabular}
\end{table}
Table~\ref{tab:full_classification_report} presents the resulting classification report for the trained model. We observe that the classification performance is relatively low when tested against all kidneys. One major reason for this outcome is the imbalanced nature of the dataset, with only about 3\% of the observations being acceptances. This imbalance can cause the model to achieve high precision without adequately learning to differentiate between classes. Additionally, differentiating between the likelihood of centers accepting a higher quality kidney is challenging, as most centers are willing to accept these kidneys when they are offered to the ``right" candidate on their waiting list. Consequently, factors such as the specific characteristics of the offer-receiving patient and the portfolio of patients listed at the center at the time of the offer play a more dominant role than center-level differences in offer acceptance behavior for higher-quality kidneys, making the proposed classification problem a highly complex one. On the other hand, it is important to note that ALP is not intended to replace the regular allocation system but rather to provide a data-driven framework to aid in intervention decisions, specifically to increase the transplantation likelihood of hard-to-place kidneys.

Despite the relatively subpar performance of the ALP framework in the classification task, its use for ranking shows significant improvement over current practices. Table~\ref{tab:acceptance} highlights the difference in the expected or observed NCS score, which represents the average number of centers that would be offered the organ before an acceptance. The baseline shows that, on average, 4.27 centers are offered the kidney before it is accepted, whereas the ALP framework reduces this to an average of 1.15 centers before reaching the accepting center. This demonstrates the potential of the ALP framework to streamline the allocation process and improve outcomes, even when applied to the allocation of all kidneys.

\begin{table}[t]
\centering
\caption{Expected/observed NCS score - uncensored data %\mycomment{S: These numbers do not match with those reported in the text. Which one is correct?}\textcolor{blue}{Fixed to match}
}
\label{tab:acceptance}
\begin{tabular}{lccc}
\toprule
Method & Expected/Observed NCS Score (avg.)\\
\midrule
Baseline & 4.273 \\
ALP Framework & 1.154 \\
KDRI Heuristic & 1.415 \\
KAP Heuristic & 3.397 \\
\bottomrule
\end{tabular}
\end{table}

%%%%%%%%%%%%%%%%%%%%%%%%%%%%%%%%%%%%%%%%%%%%%%%%%%%%%%%
\subsection{Exclusion of Low-Offer-Volume Centers in Data Censoring}
%%%%%%%%%%%%%%%%%%%%%%%%%%%%%%%%%%%%%%%%%%%%%%%%%%%%%%%

It is well documented that gradient-boosted classification models, such as LGBM, can exhibit bias towards the majority class when dealing with imbalanced data~\citep{tanha2020boosting}. To address this issue, one effective approach is to balance the training data by under-sampling the majority class. We implemented this approach by counting the number of times each center is offered a single organ and dividing this count by the total number of patients listed at that center at the time of the offer. If this ratio falls below a specified cutoff, we remove that center-organ offer observation from the dataset. 

Our goal is to exclude ``low-offer volume transplant centers'' that did not receive the organ offer for a significant fraction of their listed patients, as these centers may not be comparable to the accepting center. A transplant center that rejects an offered organ for only a small fraction, such as 5\%, of their highest priority patients might have accepted it if the offers were made to a larger fraction of their patients. By censoring the dataset, we ensure that the training data more accurately reflects scenarios where centers have a realistic chance of accepting an organ, thereby improving the ML model's ability to make precise predictions. This technique not only mitigates the imbalance issue but also enhances the interpretability and applicability of the model in real-world kidney allocation scenarios.

\begin{figure}[b]
    \centering
    \includegraphics[scale=0.5]{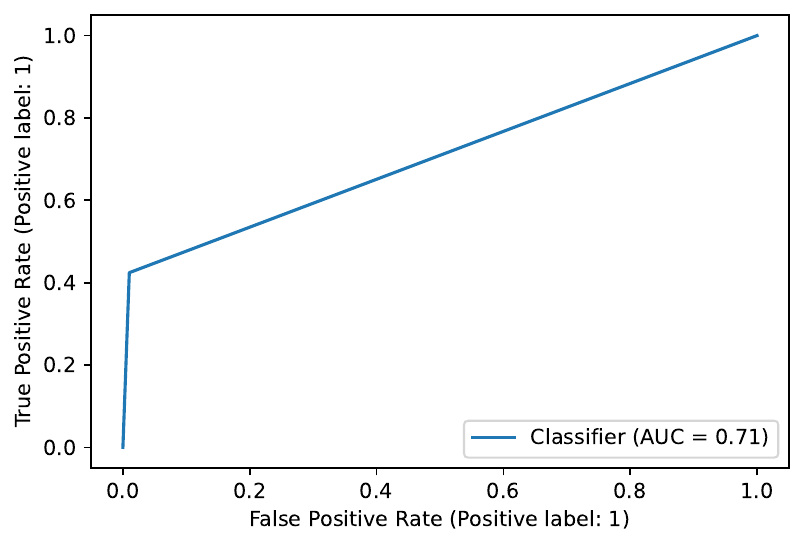}
    \caption{ROC curve for censored data with unrestricted KDRI}
    \label{fig:roc}
\end{figure}

After censoring, the dataset achieves a balance where approximately 5\% of the observations are acceptances. Table~\ref{tab:balanced_classification_report} presents the classification report for the model trained on the censored data. The accuracy for the acceptance class improves to 68.9\%, a significant enhancement compared to the performance on the uncensored data. Additionally, Figure~\ref{fig:roc} illustrates an improved ROC curve along with a higher AUC score, indicating better discrimination ability.

% It is well documented that Gradient Boosted Classification models can suffer from bias toward the majority class in the presence of imbalanced data. One approach to addressing this issue is to balance the training data by under-sampling the majority class. We have achieved this by counting the number of times a respective center is offered a single organ and dividing by the total number of their patient check-ins if this value is below a cutoff then we remove that center/organ observation. This way we are removing centers which the current allocation system is deeming less valuable.

% When the percentage of accepted observations reaches approximately 5\% of the data we achieve the following results Table~\ref{tab:balanced_classification_report} shows the classification report for the model trained on censored data, the accuracy for the acceptance class is now 68.9\% this is a significant improvement over the uncensored data. Moreover Figure~\ref{fig:roc} shows a much better ROC curve as well as AUC score. 

% This is still a very significant improvement.   

\begin{table}[!ht]
\centering
    \caption{Classification report for LGBM - censored data} % - the dataset excluding low-offer-volume centers}
\begin{tabular}{lrrrr}
\toprule
& Precision & Recall & F1 score & Support \\
\midrule
0 (Reject) & 0.971 & 0.987 & 0.979 & 173,165 \\
1 (Accept) & 0.689 & 0.495 & 0.576 & 10,136 \\
\midrule
Accuracy & 0.960 & 0.960 & 0.960 & 183,301 \\
Macro avg & 0.830 & 0.741 & 0.777 & 183,301 \\
Weighted avg & 0.955 & 0.960 & 0.957 & 183,301 \\
\bottomrule
\end{tabular}
    \label{tab:balanced_classification_report}
\end{table}

Table~\ref{tab:acceptance_bal} presents the expected or observed NCS scores after applying data censoring. The baseline shows an average of 1.97 centers offered before the accepting center receives an offer, while the ALP framework reduces this to just 0.36 centers on average. This represents a significant improvement, demonstrating the effectiveness of the data censoring approach in enhancing the performance of the ALP framework.

\begin{table}[ht]
\centering
\caption{Expected/observed NCS score - censored data}
\label{tab:acceptance_bal}
\begin{tabular}{lccc}
\toprule
Method & Expected/Observed NCS Score (avg.) \\
\midrule
Baseline & 1.97 \\
ALP Framework & 0.36 \\
\bottomrule
\end{tabular}
\end{table}

%%%%%%%%%%%%%%%%%%%%%%%%%%%%%%%%%%%%%%%%%%%%%%%%%%%%%%%
\subsection{Performance of the ALP Framework over Hard-to-Place Kidneys: KDRI Range Restriction}
%%%%%%%%%%%%%%%%%%%%%%%%%%%%%%%%%%%%%%%%%%%%%%%%%%%%%%%
KDRI is a widely used metric that reflects the relative risk of post-transplant kidney graft failure from a specific deceased donor compared to a median reference donor~\citep{park2019association}. Lower KDRI scores are associated with longer estimated graft function, while higher KDRI scores are associated with shorter estimated function. For example, a kidney with a KDRI of 80\% is expected to have shorter longevity than 80\% of recovered kidneys. This detailed metric considers several donor factors including age, height, weight, ethnicity, history of hypertension or diabetes, cause of death, serum creatinine, hepatitis C virus status, and donation after cardiac death status.

In our previous analyses (see \Cref{sc:ALP_performance_unrestricted}), we evaluate the performance of the ALP framework using the entire dataset without any restrictions on the KDRI range. However, as previously discussed, most transplant centers are willing to accept lower KDRI (higher quality), e.g., KDRI $\le$ 1.3, kidneys for at least some of their listed patients. Therefore, a center's likelihood of accepting a lower KDRI kidney is primarily influenced by dynamic, patient-specific factors rather than the center's overall offer acceptance characteristics. Thus, predicting a transplant center's acceptance likelihood of a lower KDRI kidney using empirical data is challenging due to the high variability in patient-specific factors, as illustrated by the model performances in \Cref{sc:ALP_performance_unrestricted}. Additionally, since lower KDRI kidneys generally do not face significant risks of going unused, organ allocation interventions primarily target hard-to-place kidneys with relatively higher KDRI values. These higher KDRI kidneys are more challenging to allocate and are at a greater risk of non-use, making them the focus of efforts to improve organ utilization and reduce non-use rates.

Although there is no consensus on the definition of hard-to-place kidneys in the literature, data indicates that kidneys with high KDRI values face a significantly higher risk of non-use compared to those with a lower KDRI~\citep{mohan2018factors}. Accordingly, we refer to the set of kidneys with KDRI between 1.65 and 2 as hard-to-place kidneys, which are expected to be the primary targets of organ allocation interventions. We repeat the experiments previously outlined for this KDRI-restricted set of kidneys. Note that we use an upper bound of 2 on the KDRI of the selected set of kidneys because organs with KDRI $>$ 2 suffer from extreme non-use rates. The lack of accepted kidneys in this KDRI range interferes with ML model learning and makes the already imbalanced prediction task even more challenging.

\begin{table}[!ht]
\centering
    \caption{Classification report for LGBM - hard-to-place kidneys-censored data}
\begin{tabular}{lrrrr}
\toprule
& Precision & Recall & F1 score & Support \\
\midrule
0 (Reject) & 0.98 & 0.99 & 0.99 & 20,770 \\
1 (Accept) & 0.82 & 0.55 & 0.66 & 1,063 \\
\midrule
Accuracy & 0.97 & 0.97 & 0.97 & 21,833 \\
Macro avg & 0.90 & 0.77 & 0.82 & 21,833 \\
Weighted avg & 0.97 & 0.97 & 0.97 & 21,833 \\
\bottomrule
\end{tabular}

    \label{tab:class1.65}
\end{table}

\begin{figure}[t]
    \centering
    \includegraphics[scale=0.5]{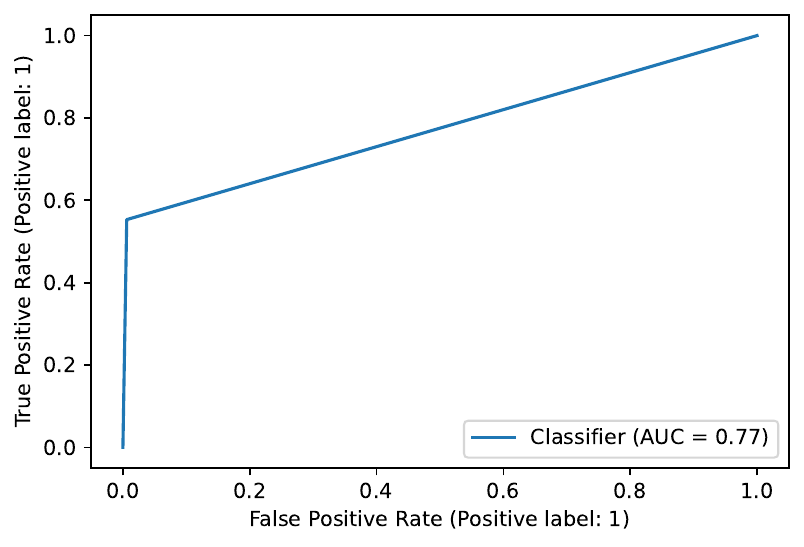}
    \caption{ROC curve for hard-to-place kidneys - censored data}
    \label{fig:roc1.65}
\end{figure}

Table~\ref{tab:class1.65} shows the classification report summarizing the performance of the ALP framework in predicting the likelihood of transplant centers accepting hard-to-place kidneys, with the dataset balanced similarly to the previous analysis. The classification report indicates very high accuracy for both acceptance and rejection classes, suggesting that our ranking system is a robust choice. Table~\ref{tab:acceptance_bal1.65} presents the average expected or observed NCS score for both the baseline and the ALP framework. Despite both values being below 1 due to censoring, the NCS score of the ALP framework significantly outperforms the baseline by a factor of 10. Finally, \Cref{fig:roc1.65} presents the resulting ROC curve for predicting transplant centers' likelihood of accepting hard-to-place organs, indicating a slight improvement in performance compared to the analysis using the unrestricted data (see \Cref{fig:roc}). This improvement suggests that the ALP framework is more effective in predicting the acceptance likelihood for hard-to-place kidneys within the specified KDRI range, thereby improving the accuracy and reliability of the ALP framework within the organ allocation process.

% \begin{table}[ht]
% \centering
% \begin{tabular}{lccc}
% \toprule
% Method & Expected/Observed Number of No's \\
%  & Before an Acceptance (AVG) \\
% \midrule
% Baseline & 0.693 \\
% C-Rank & 0.028 \\
% \bottomrule
% \end{tabular}
% \caption{Expected/Observed Number of No's Before an Acceptance}
% \label{tab:acceptance_bal1.65}
% \end{table}

\begin{table}[ht]
\centering
\caption{Expected/observed NCS score for hard-to-place kidneys - censored data}
\begin{tabular}{lc}
\toprule
Method & Expected/Observed NCS score (avg.)\\
\toprule
Baseline & 0.463 \\
ALP Framework & 0.045 \\
% Average Heuristic & 0.134 \\
KDRI-based Heuristic & 0.124 \\
KAP-based Heuristic & 0.441 \\
\bottomrule
\end{tabular}

\label{tab:acceptance_bal1.65}
\end{table}

These findings highlight the effectiveness of the ALP framework in improving the allocation process for hard-to-place kidneys. By focusing on kidneys within the KDRI range of 1.65 to 2, the framework effectively identifies transplant centers with a higher likelihood of acceptance, thereby reducing the number of rejections and the resulting CIT. This targeted approach has the potential to not only enhance the timeliness and efficiency of the organ allocation system but also maximize the utilization of available kidneys, ultimately improving patient outcomes.

% \begin{figure}[!ht]
%  \centering
%  \vspace*{5pt}%
%  \hspace*{\fill}% 
%   \begin{subfigure}{0.8\textwidth}     % start subfigure 1
%     %\belowcaptionskip=8pt
%     \centering
%     \includegraphics[width=\textwidth]{figures/local/local2high.png}%
%     \captionsetup{skip=12pt}%
%     \caption{Impact of features on acceptance probability top-ranked center}
%     \label{fig:demo3}
%   \end{subfigure}%          % end subfigure 1
%   \hspace*{\fill}%          % empty line absolutely necessary!

%   \vspace*{8pt}%  

%   \hspace*{\fill}%  
%    \begin{subfigure}{0.8\textwidth}        % start subfigure 2
%     %\belowcaptionskip=8pt
%     \centering
%     \includegraphics[width=\textwidth]{figures/local/local2low.png}%
%     \captionsetup{skip=12pt}%
%     \caption{Impact of features on acceptance probability bottom ranked center}
%     \label{fig:demo4}
%   \end{subfigure}%          % end subfigure 2  
%   \hspace*{\fill}%          % empty line absolutely necessary!
%   % \captionsetup{skip=8pt, format=nocap}% % required to hide the figure (main) lable!
%   \caption{Local Interpretation of the model acceptance probability}
%   \label{fig:loc2}
% \end{figure}

%%%%%%%%%%%%%%%%%%%%%%%%%%%%%%%%%%%%%%%%%%%%%%%%%%
\subsection{Interpretability}
%%%%%%%%%%%%%%%%%%%%%%%%%%%%%%%%%%%%%%%%%%%%%%%%%%

\begin{figure}
    \centering
    \includegraphics[width=.8\textwidth]{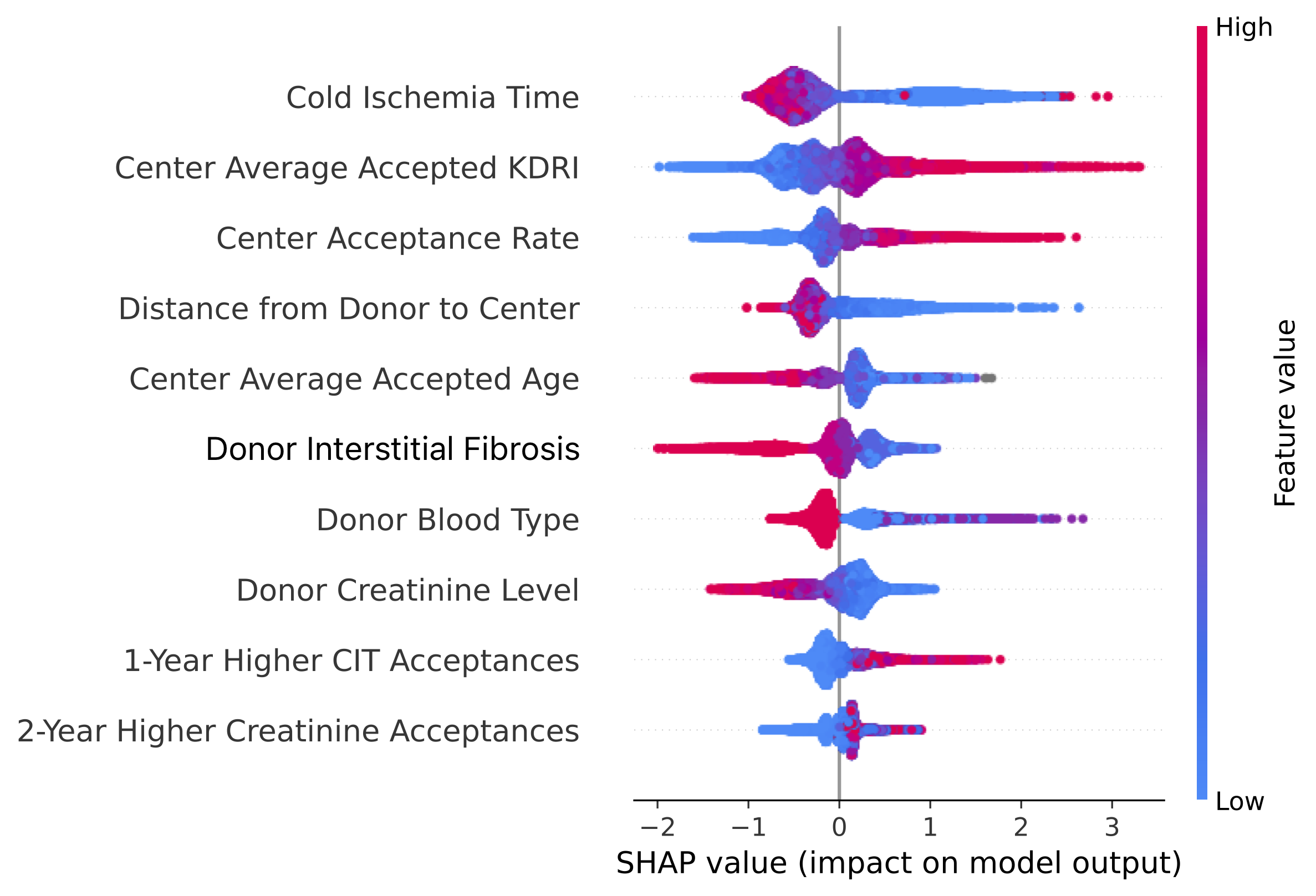}
    \caption{SHAP beeswarm plot for high KDRI kidneys (showing top 10 features)}
    \label{fig:shapforce165}
\end{figure}

% Interpretability is a critical aspect of deploying ML models, particularly in sensitive and high-impact domains such as organ allocation. It ensures that the decision-making process of the models is transparent, understandable, and justifiable to all stakeholders, including medical professionals and patients. By providing clear insights into how predictions are made, interpretability fosters trust, facilitates model validation, and aids in identifying potential biases~\citep{doshi2017towards}. 
In this section, we discuss the methods used to enhance the interpretability of our ALP framework, ensuring that its predictions are not only accurate but also transparent and actionable.
\Cref{fig:shapforce165} displays the SHAP beeswarm plot for the ALP framework predicting the centers' transplantation likelihood of hard-to-place kidneys. A beeswarm plot provides an information-rich summary of how the most significant features in a dataset influence the model's output. Each instance in the dataset is represented by a single dot on the respective feature row. The x-axis position of each dot corresponds to the SHAP value of that feature, and dots accumulate along each row to indicate density. The color of the dots represents the original value of the feature, adding an additional layer of insight into the feature's impact. 

% \mycomment{S: A few notes. First, I believe this plot is called beeswarm, and not the force plot, see \url{https://shap.readthedocs.io/en/latest/example_notebooks/api_examples/plots/beeswarm.html}. Please correct the captions accordingly. Second, for beeswarm plots, reporting "Sum of X other features", where X is the number of remaining variables, might be nice. In any case, we should indicate in the caption that these are the top Y most significant features. Third, the two beeswarm figures we provide should be consistent, including the number of features we are reporting and the labeling of the features, the format of \Cref{fig:shapforce165} is great, please use the same for the other figure.}
 
% The SHAP beeswarm plot illustrated in \Cref{fig:shapforce165} visually represents the impact of features on the model's predictions. Each dot represents an individual prediction, with color indicating the feature’s value (blue for low, red for high), and the position along the X-axis shows the feature's impact (right for positive, left for negative). 
In \Cref{fig:shapforce165}, we observe that CIT emerges as the most impactful feature on the model's output, where higher CIT values significantly reduce the likelihood of acceptance, while lower CIT values increase it. This finding aligns with clinical understanding, as prolonged ischemia time is known to adversely affect kidney viability. Next in importance are the center-specific average accepted KDRI and the center's average acceptance rate. Both of these factors show a positive correlation with acceptance probability, indicating that transplant centers with a track record of accepting higher KDRI kidneys and those with higher overall acceptance rates are more likely to accept the offered kidney. This may suggest that such centers may have developed expertise and protocols to manage higher-risk kidneys effectively. Alternatively, this could suggest that centers with higher volumes of transplants may be able to accept an allograft failure without hurting their outcomes states. Additionally, the distance from the donor hospital to the center and the center's average accepted age display a negative correlation with acceptance probability. Greater distances likely introduce logistical challenges and potential delays, reducing the chances of acceptance. Similarly, centers that tend to accept younger donor kidneys may be more selective, thereby lowering the acceptance likelihood for hard-to-place kidneys.

Several donor-specific features, including blood type and donor creatinine levels, also play significant roles. Compatible blood types and lower creatinine levels, indicating better kidney function, positively influence acceptance probability and serve as distinguishing donor-related features among hard-to-place kidneys. Moreover, dynamic features such as the number of kidneys a center accepted in the past year with higher CIT and the number of kidneys accepted in the past two years with higher creatinine levels, indicate that centers familiar with managing kidneys under less favorable conditions are more likely to accept similar kidneys again.

% \mycomment{S: For \Cref{fig:double_image}, we need to specify what we mean by ``average" in this context; whether it refers to average KDRI, CIT, or other characteristics. We may include a short table in the Appendix to report the specific values for this kidney (at least key features), similarly for \Cref{fig:double_image2}.\textcolor{blue}{I have added these tables  to the appendix and referenced them in the fig caption}}

\begin{figure}[!htp]
    \centering
    \begin{subfigure}{.72\textwidth}
        \includegraphics[width=\textwidth]{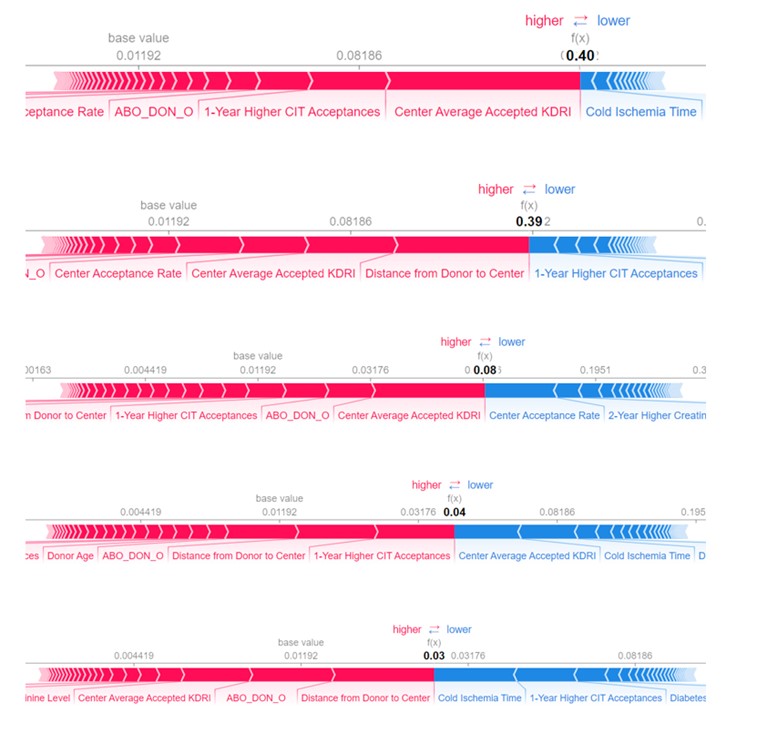}
        \caption{Top five centers most likely to accept}
    \end{subfigure}%

    \begin{subfigure}{.72\textwidth}
        \includegraphics[width=\textwidth]{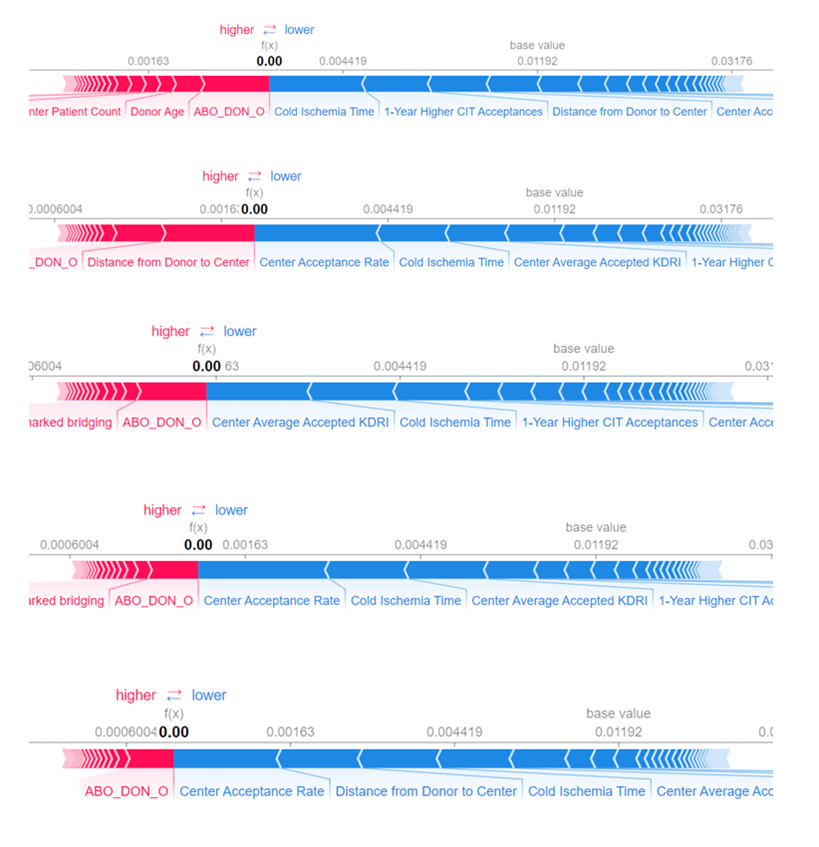}
        \caption{Top five centers least likely to accept}
    \end{subfigure}
    \caption{SHAP local force plots for the top and bottom five centers ranked by likelihood of accepting an average hard-to-place kidney; highlighting the most influential features in the classification. Summary stats for this kidney can be found in \Cref{num1}.
    }
    \label{fig:double_image}
\end{figure}

\begin{figure}[!htp]
    \centering
    \begin{subfigure}[b]{.72\textwidth}
        \includegraphics[width=\textwidth]{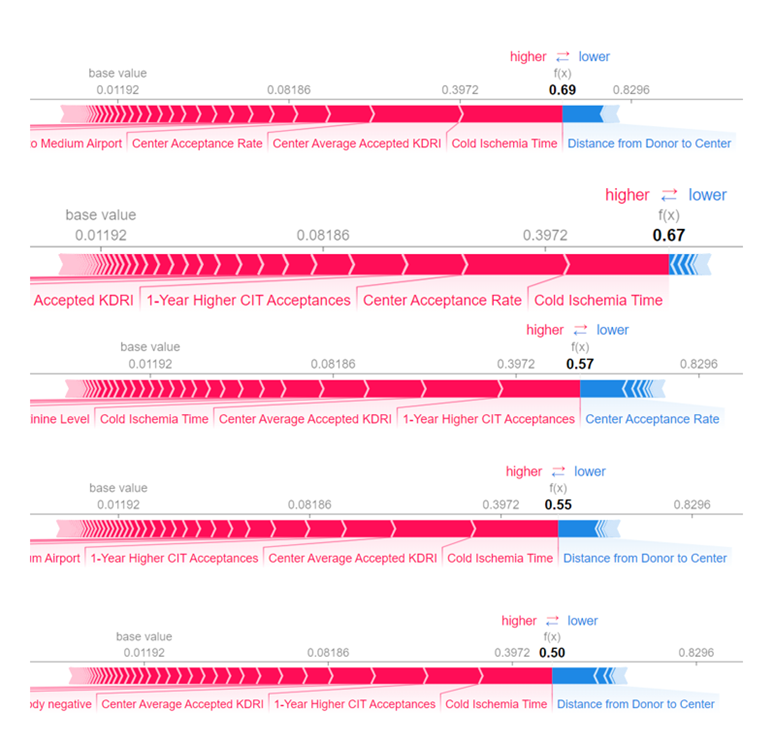}
        \caption{Top five centers most likely to accept %\mycomment{The font size in this one is really good. Can we do the same thing for the other three?}
        }

    \end{subfigure}
    
    \begin{subfigure}[b]{.72\textwidth}
        \includegraphics[width=\textwidth]{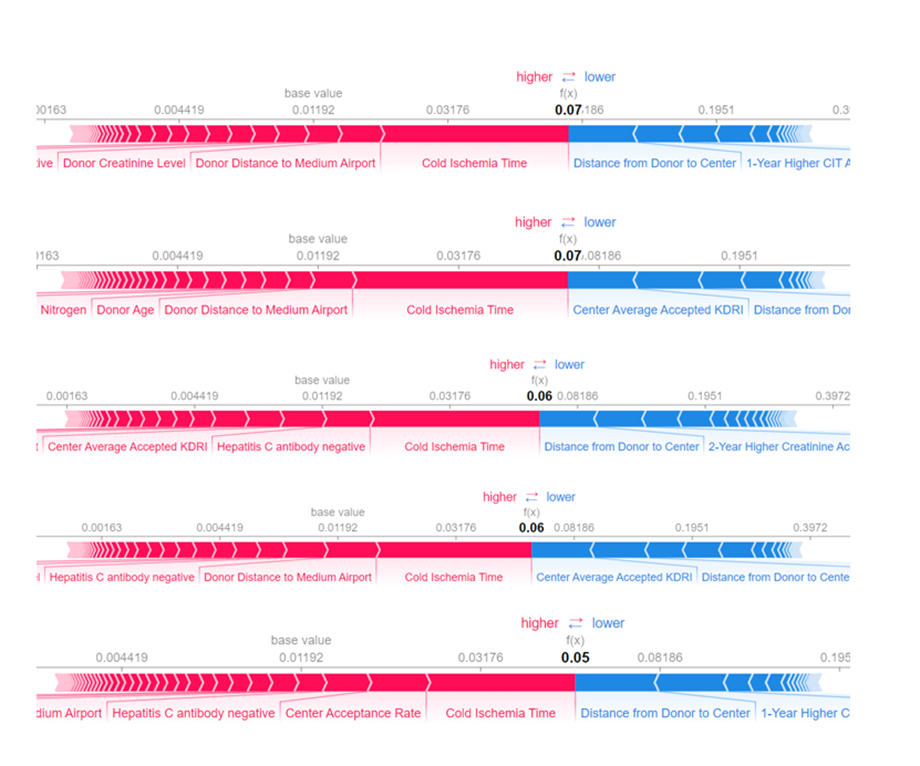}
        \caption{Top five centers least likely to accept}

    \end{subfigure}
    \caption{SHAP local force plots for the top and bottom five centers ranked by likelihood of accepting a hard-to-place kidney with very low CIT, very close to large airport, with a history of smoking; highlighting the most influential features in the classification. Summary stats for this kidney can be found in~\Cref{num2}.}
    \label{fig:double_image2}
\end{figure}

\Cref{fig:double_image,fig:double_image2} illustrate the key factors influencing the ALP's estimated probability of acceptance for the top and bottom five centers ranked by their likelihood of accepting two distinct hard-to-place kidneys. The first kidney (\Cref{fig:double_image}) represents an average hard-to-place kidney, denoting a KDRI in the median for the 1.65 to 2.0 range, with a KDRI of 1.824 and an initial IT at time of first offer of 594 minutes. %\textcolor{blue}{ (Cold ischemia time is dynamic e.g., each offer has a different cit I have just put the first one here)}. 
For this kidney, the features contributing most to the acceptance likelihood for the top five centers likely to accept are predominantly center-related behaviors. These include the average KDRI of accepted kidneys, the distance from the donor to the center, as well as the center's overall kidney offer acceptance rate. Additionally, factors such as the number of higher CIT kidneys a center has accepted in the past year as well as the blood type and age of the donor themselves. Conversely, features contributing to the rejection likelihood include the centers' relative lack of higher CIT acceptances within the past year, the kidney's high accumulated CIT as well as the donor's history of diabetes. Despite these rejection-indicating features, the centers' historical acceptance behavior of similar kidneys is more influential, leading the model to predict a high likelihood of acceptance for these top centers.

% These include the average KDRI and age of kidneys accepted by the center in the past two years, as well as the center's overall offer acceptance rate. Additionally, factors such as the center's proximity to airports and the nominal GDP of the center's state reinforce the acceptance likelihood. Conversely, features contributing to the rejection likelihood include the centers' relative lack of higher CIT acceptances within the past year, the donor's history of myocardial infarction (MI), and the donor's blood type. Despite these rejection-indicating features, the centers' historical acceptance behavior of similar kidneys was more influential, leading the model to predict a high likelihood of acceptance for these top centers. 

In contrast, for the bottom five centers that are least likely to accept, the same center-related behaviors, including the average KDRI and age of accepted kidneys and the overall offer acceptance rate are significant factors contributing to the rejection likelihood. Although other factors, such as the lack of interstitial fibrosis on the kidney, the donor's blood type and age as well as proximity to the center positively influence the acceptance likelihood, these centers' historical behavior towards similar hard-to-place kidneys result in a higher predicted likelihood of rejection.

The second kidney (\Cref{fig:double_image2}) represents a hard-to-place kidney with very low IT, with the first several offers having a zero CIT, originating from a hospital in close proximity to a medium airport, a medium airport being an airport accepting more than 0.25\%, but less than 1\% of annual passenger boardings nationwide. For this kidney, similar to the first one, center-related behaviours are among the most influential features contributing to the acceptance likelihood for the top five centers that are likely to accept. However, the impact of the very low CIT predominantly outweigh other factors. The proximity of the donor hospital to a medium airport also plays a significant role, making the centers' closeness to medium and large airports more pertinent in this scenario compared to the previous kidney. Despite the donor's seemingly large distance to several centers being a factor contributing to rejection, the extremely low CIT and the centers' historical acceptance behaviour lead the model to predict a high likelihood of acceptance for these centers.

For the bottom five centers that are least likely to accept, the pattern is again similar to that observed in \Cref{fig:double_image}. Center-related behaviours are significant factors contributing to the rejection likelihood. Although the very low CIT is a significant factor suggesting acceptance, its impact, along with other positive features such as low creatinine levels and the centers' proximity to the donor hospital, is overshadowed by the centers' historical lack of acceptance for similar hard-to-place kidneys.

% \mycomment{S: I think there are no significant new insights presented in \Cref{fig:shapforce,fig:loc1} compared to the previous analysis. Also, since interpretability is more relevant for the practical setting of predicting the acceptance likelihood of hard-to-place kidneys, these results for all kidneys disrupt the flow of the main text. So, I moved these figures to the Appendix for better coherence. Bring them back if you disagree.}

%%%%%%%%%%%%%%%%%%%%%%%%%%%%%%%%%%%%%%%%%%%%%%%%%%%%%%%%%%%%%%%%%%%%%%%%%%%%%%%%%

%%%%%%%%%%%%%%%%%%%%%%%%%%%%%%%%%%%%%%%%%%%%%%%%%%%%%%%%%%%%%%%%%%%%%%%%%%%%%%%%%
\section{Conclusion}
\label{conc}
%%%%%%%%%%%%%%%%%%%%%%%%%%%%%%%%%%%%%%%%%%%%%%%%%%%%%%%%%%%%%%%%%%%%%%%%%%%%%%%%%
The increasing demand for kidney transplants, coupled with persistent organ scarcity, necessitates the development of more efficient and equitable allocation systems. Hard-to-place kidneys, in particular, pose a critical challenge, often leading to high non-utilization rates and missed opportunities to save lives. Our research addresses this urgent need by introducing a data-driven, ML-based approach that optimizes and expedites the allocation of hard-to-place kidneys to transplant centers with the highest likelihood of acceptance.

Our study distinguishes itself by focusing on the optimization of out-of-sequence offers and the allocation of hard-to-place kidneys—an area that has received limited attention in the literature. We propose a novel ML-based ranking system that incorporates a comprehensive set of features, many of which have been underexplored in previous studies. These features include dynamic calculations of CIT at the time of offer, center-specific acceptance behaviors, and logistical factors such as the proximity of transplant centers and donor hospitals to airports. This holistic approach significantly improves the model's ability to predict which centers are most likely to accept hard-to-place kidneys and provides a powerful ranking mechanism to enhance the overall allocation process.

The results from our detailed numerical study demonstrate the effectiveness of the proposed model and illustrate its superiority over traditional methods and existing heuristics. Our ML-based ranking system shows a substantial improvement in reducing the average number of rejections before a kidney is accepted, decreasing this number by a factor of 16 compared to the existing allocation system. This marked improvement highlights the potential of our approach to streamline the allocation process, reduce CIT, and ultimately increase the utilization of hard-to-place kidneys and improve their post-transplant utility by expediting their allocation. Additionally, our research introduces an innovative evaluation framework that allows for a rigorous comparison of our model against existing heuristics and baseline practices, further establishing its effectiveness across various scenarios, including different KDRI ranges.

%\mycomment{S: What do we mean by this 84 minutes delay. I think this is an important discussion point, but I wanted to clarify before elaborating on this finding.} 
Our analysis also uncovers a concerning average delay of 84 minutes between each center's responses for hard-to-place kidneys. Given the critical impact of CIT on transplant outcomes, with each passing hour significantly increasing the risk of graft failure and patient mortality~\citep{debout2015each}, this delay is far from trivial. However, our experiments demonstrate that the proposed placement policy can reduce the average number of centers considered before placement by fourfold for all kidneys and tenfold for hard-to-place kidneys. This substantial reduction in the number of rejections before acceptance suggests that our method not only expedites the acceptance process but also improves the overall utilization of hard-to-place kidneys. By minimizing delays and enhancing the speed of kidney placement, our approach has the potential to significantly decrease the risks of short- and long-term graft failure, ultimately leading to better transplant outcomes.

In addition to the predictive accuracy, our study emphasizes the importance of model interpretability, which is crucial for the clinical adoption of such systems. We conducted a detailed analysis using both local and global interpretability methods, offering insights into the most influential features driving the model’s predictions. This transparency not only fosters trust in the system but also offers valuable guidance for clinicians and policymakers, allowing them to refine kidney allocation and placement strategies based on data-driven insights. Our findings reveal the critical importance of center-specific features and demonstrate how the interplay of these factors can be leveraged to optimize the placement of hard-to-place kidneys.

While our findings are promising, we recognize certain limitations that must be addressed before real-world implementation. One critical concern is that centers with a historically low volume or limited experience in accepting hard-to-place kidneys may struggle to receive such offers over time, even if they aim to become more aggressive in their transplant practices. This raises the issue of fairness in the allocation system. To address this, we advocate for regularly updating the model to reflect changes in centers' behavior and acceptance patterns. By capturing new trends in acceptance behavior, the model can adapt to centers that are willing to take on higher-risk transplants. Additionally, ensuring that only truly hard-to-place organs are subjected to out-of-sequence allocations would limit the scope of this policy, giving historically lower-volume centers the opportunity to demonstrate their willingness and capacity to accept such organs. We also emphasize that the real-world deployment of this model would require regulatory approval, rigorous validation, and careful consideration of patient safety and ethical implications. These steps are essential to ensure that the model can be safely and effectively integrated into the organ allocation system, without introducing unintended consequences or exacerbating existing disparities.

Looking ahead, future research could extend this work by developing a temporal-based model that accounts for past offers, acceptances, and rejections at each center when ranking them for allocation. This approach would acknowledge that the suitability of a kidney for a particular center is not a static decision made in isolation but a dynamic and evolving one influenced by past decisions and outcomes, providing a more nuanced and context-aware approach to organ allocation decision-making. By incorporating this temporal context, we can further enhance the precision of the model and uncover deeper insights into the complex factors driving kidney allocation in real-world scenarios.

% \revnote{In conclusion, 
In conclusion, our research demonstrates the substantial potential of machine learning to transform kidney allocation processes. While our current analysis is focused specifically on kidney allocation, which suffers from particularly high nonuse rates, the methodology we propose can be easily generalized to other organ allocation systems. By addressing key inefficiencies and offering a robust, data-driven framework, this study not only improves kidney allocation but also paves the way for more sophisticated, efficient, and equitable intervention mechanisms and expedited placements across the broader spectrum of organ allocation. Our approach has the potential to reduce nonuse rates, optimize the utilization of hard-to-place organs, and ultimately improve patient outcomes across various types of organ transplants.%}
%%%%%%%%%%%%%%%%%%%%%%%%%%%%%%%%%%%%%%%%%%%%%%%%%%%%%%%%%%%%%%%%%%%%%%%%%%%%%%%%%
\bibliographystyle{abbrvnat} % abbrv, abbrvnat, plainnat
\bibliography{Ref}
\newpage
\appendix

\setcounter{page}{1}
\renewcommand{\thepage}{S-\arabic{page}}
\setcounter{figure}{0}
\renewcommand{\thefigure}{A\arabic{figure}}
\setcounter{table}{0}
\renewcommand{\thetable}{A\arabic{table}}

\appendix
\section{Appendix}

\subsection{Sensitivity-Specificity C-Rank Plots}
The sensitivity-specificity C-Rank plots shown in Figure~\ref{fig:ss_full}, and~\ref{fig:ss_dom} are generated as follows: 
\begin{itemize}\setItemSep{0.3em}
    \item Assembled the dataset containing model predictions, true labels, and other relevant metadata.
    \item Applied varying levels of data censoring on the training data.
    \item For each model and censoring level, calculated sensitivity, specificity, and rank where applicable across 100 different decision thresholds.
    \item Plotted sensitivity and specificity on the y-axis for all models and decision thresholds.
    \item Generated non-dominated plots by filtering out data points where another model had both higher sensitivity and specificity at the same threshold.
\end{itemize}

\subsection{Global Importance in Segmented Kidney Categories}
\label{app:glob}

\begin{figure}[!ht]
    \centering
    \includegraphics[width=\linewidth]{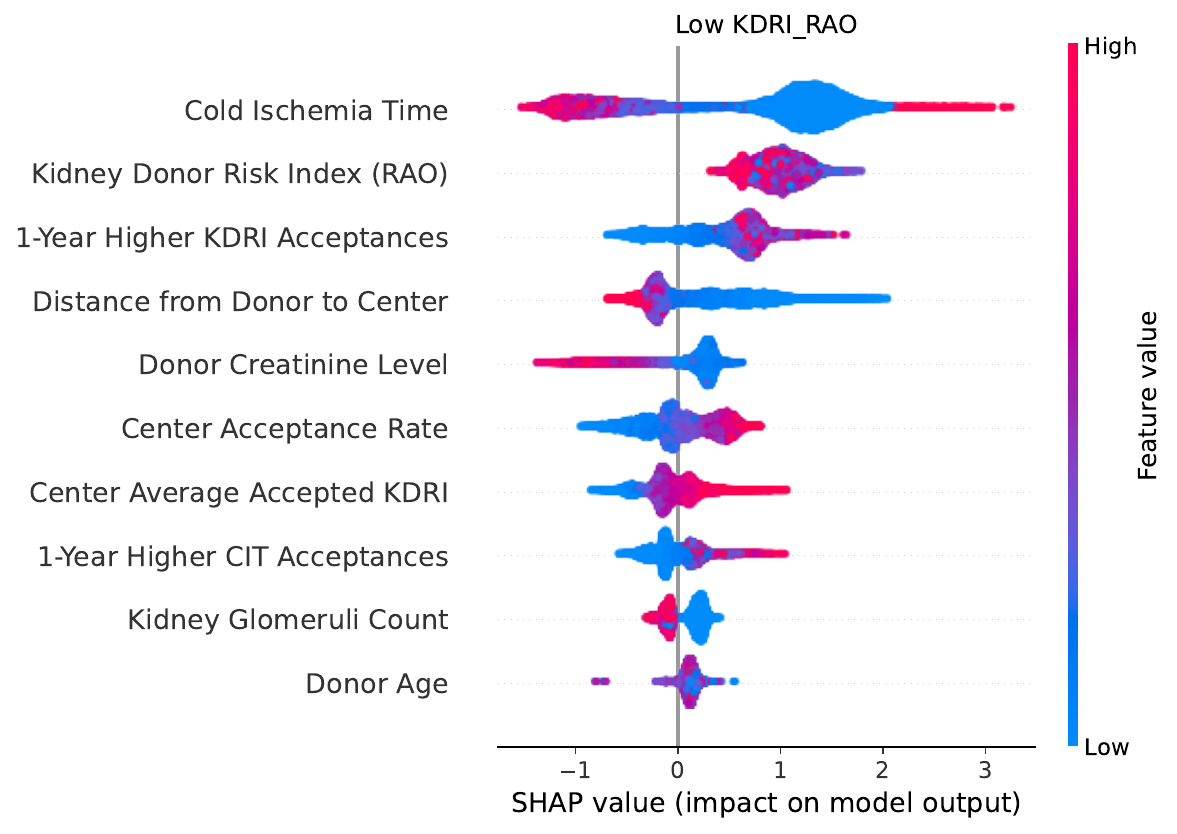}
    \caption{Global Importance Bottom 10\% of KDRI}
    \label{fig:enter-label}
\end{figure}

Figure~\ref{fig:enter-label} shows that CIT is the most dominant feature and also having the most spread. This suggests that since all kidneys in this range are ``good'' as indicated by KDRI, the model differentiates mostly with CIT. Despite being restricted to the bottom 10\% of KDRI, there are still strong variations in the SHAP analysis which demonstrates the power of this feature. Also note that since all of these are high quality, the model suggests that both a high-value or a low value in this range positively impact acceptance probability. 

\begin{figure}[!ht]
    \centering
    \includegraphics[width=\linewidth]{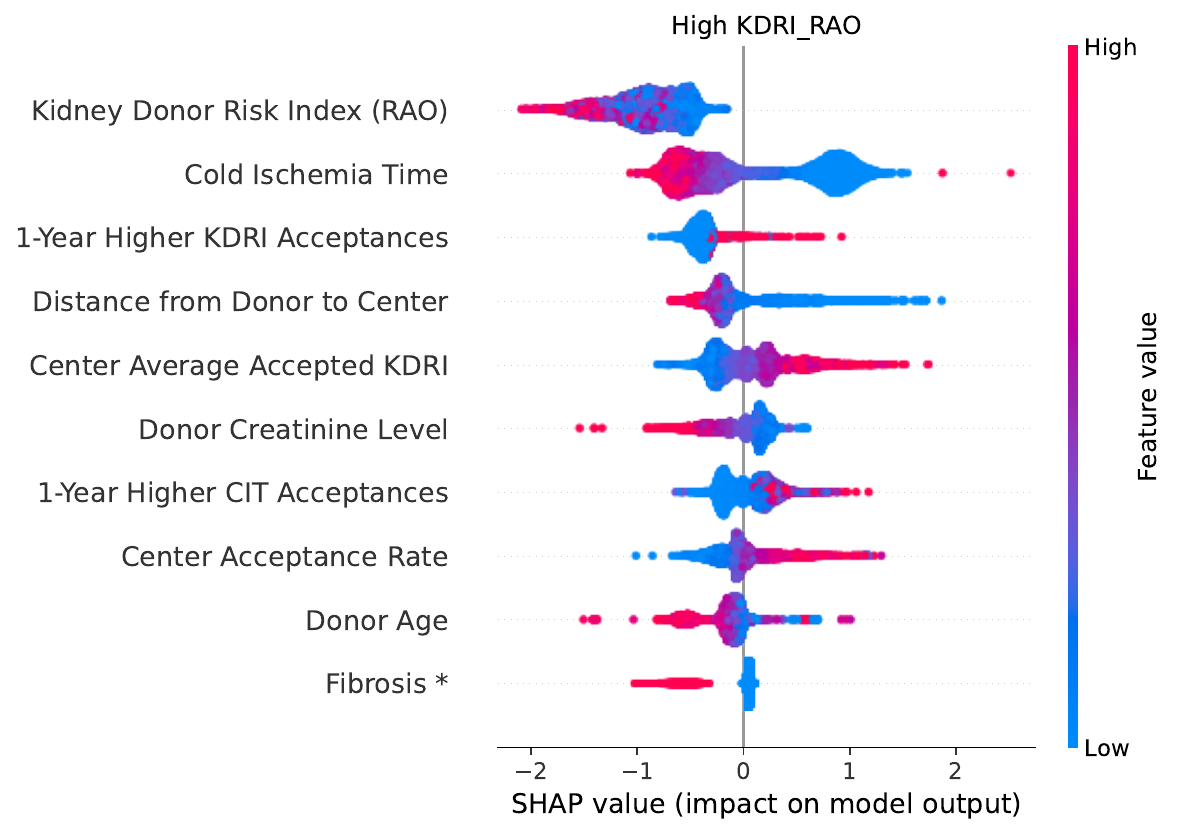}
    \caption{Global Importance Top 10\% of KDRI}
    \label{fig:enter-label1}
\end{figure}

Figure~\ref{fig:enter-label1} shows that KDRI always have a reducing effect on the acceptance probability for these kidneys. This is expected since these kidneys have high KDRI values. We see that, similar to the bottom 10\% of KDRI case, CIT shows significant variance suggesting that this feature is important for determining acceptance. Furthermore, due to the lower quality nature of these kidneys, we require the CIT to be very low to be considered. Therefore, center average accepted KDRI becomes a much more important feature for this data since some centres seem more picky than the others. 

\begin{figure}[!ht]
    \centering
    \includegraphics[width=\linewidth]{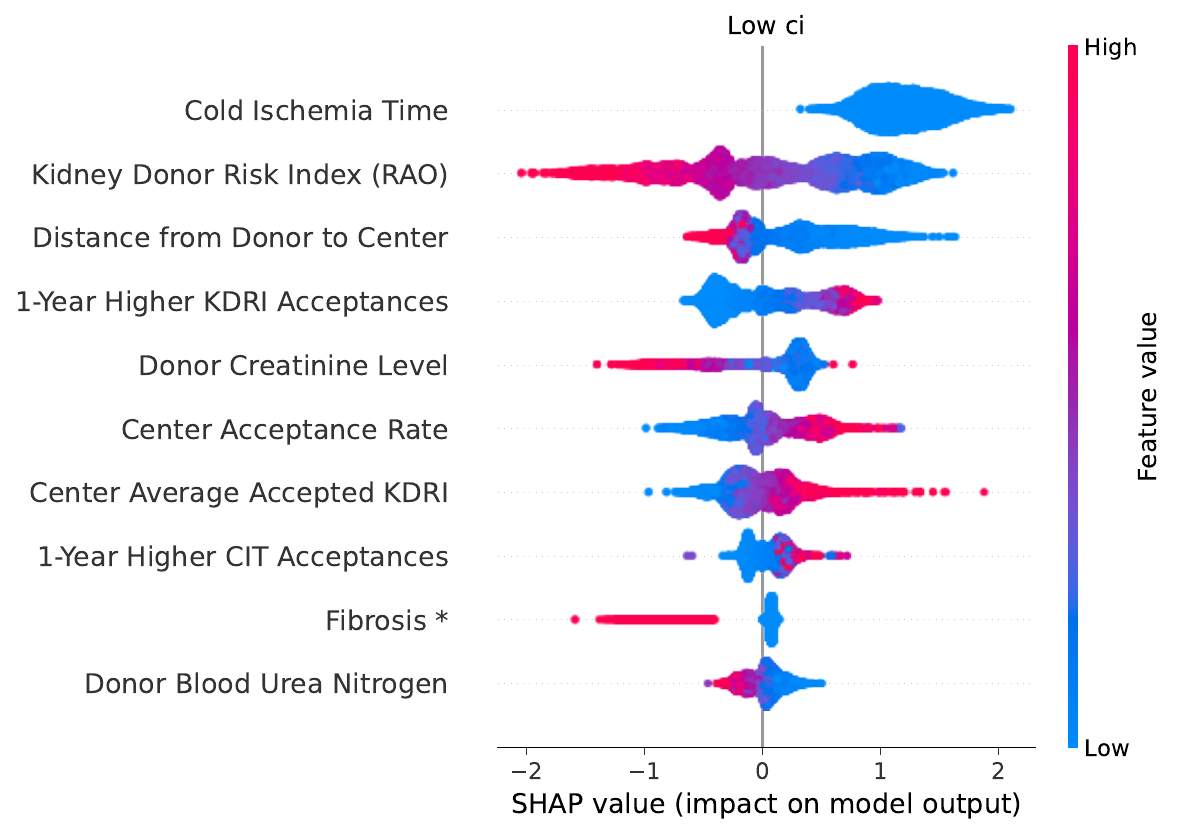}
    \caption{Global Importance Bottom 10\% of CIT}
    \label{fig:enter-label2}
\end{figure}

Figure~\ref{fig:enter-label2} shows that CIT is the most dominant feature. We note that since this graphic pertains to the lowest of CITs only, any value in this range is a positive on acceptance probability. Once again we also see that when we restrict an important feature to a subset, another feature must take over. Here we see that KDRI is bearing the weight of determining high-value kidneys. Here we also see that distance from donor to center becomes more important for the determination. 

\begin{figure}[!ht]
    \centering
    \includegraphics[width=\linewidth]{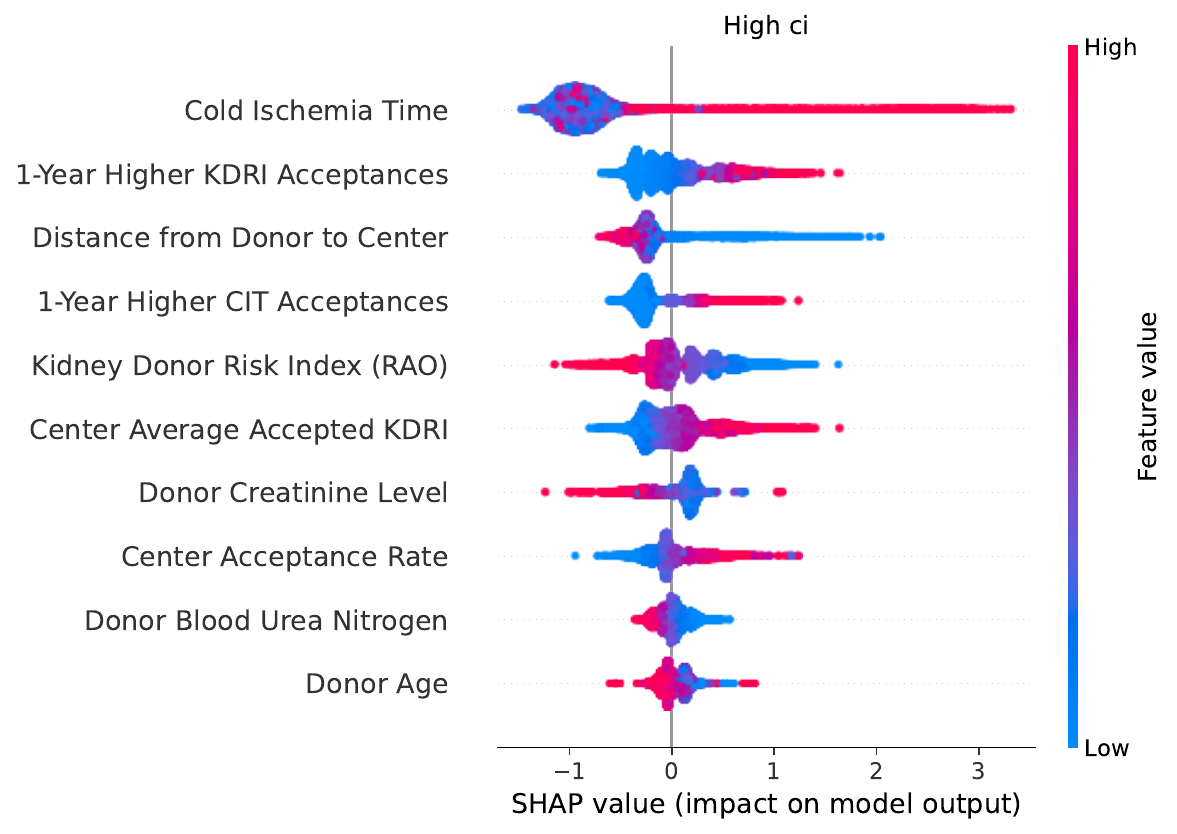}
    \caption{Global Importance Top 10\% of CIT}
    \label{fig:enter-label3}
\end{figure}

Figure~\ref{fig:enter-label3} shows that, when focusing on bottom $10\%$ CIT, higher CITs are associated with higher acceptances. This may be the result of interventions for kidneys that are not going to be used. As the next strongest feature, we see the 1-year Higher KDRI Acceptances. This suggests that when CIT is high, we look to centers that traditionally take higher volumes of lower-quality kidneys. We also observe distance from donor to center as another important feature, particularly for high $CIT$ kidneys since the risk of graft failure increases dramatically with time.

\begin{figure}[!ht]
    \centering
    \includegraphics[width=\linewidth]{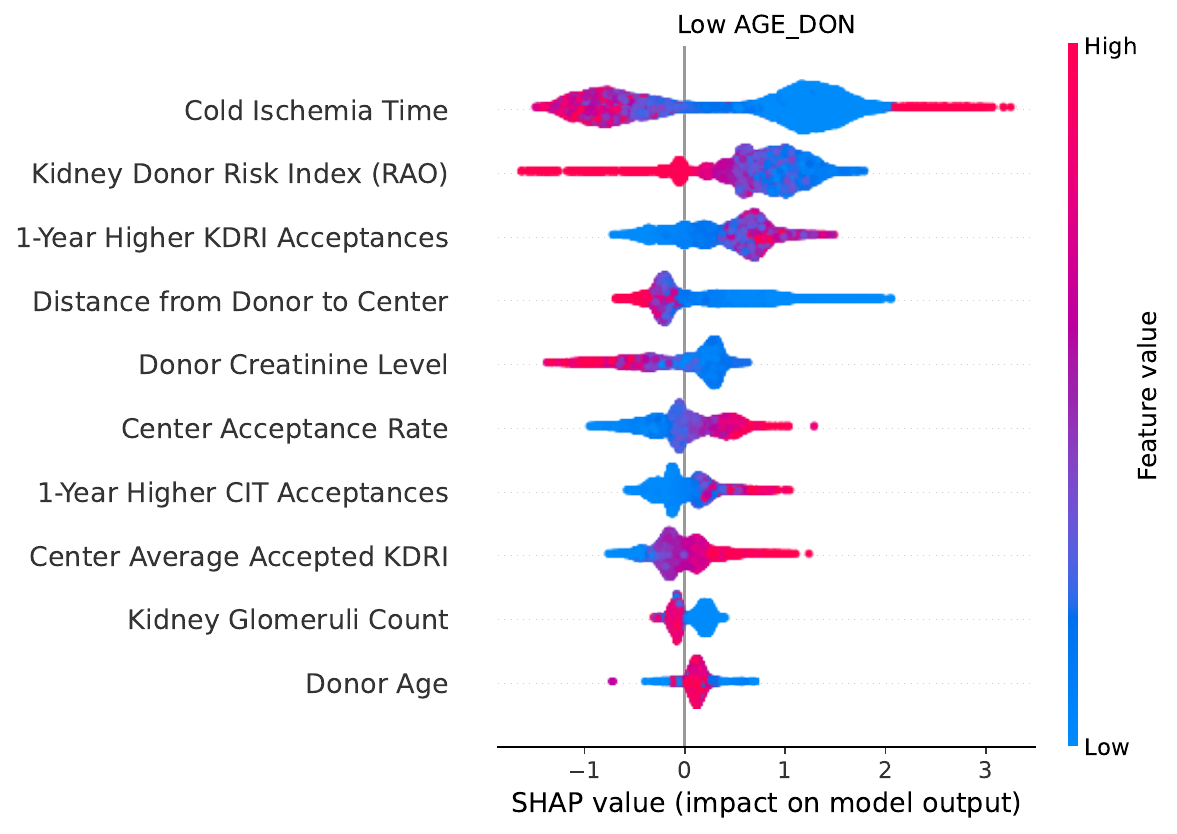}
    \caption{Global Importance Bottom 10\% of Age}
    \label{fig:enter-label4}
\end{figure}

Figure~\ref{fig:enter-label4} illustrates the continued importance of KDRI in the context of very young donors. Since KDRI takes into account age in its calculation, this implies that there is still significant variation in KDRI even at this age. 

\begin{figure}[!ht]
    \centering
    \includegraphics[width=\linewidth]{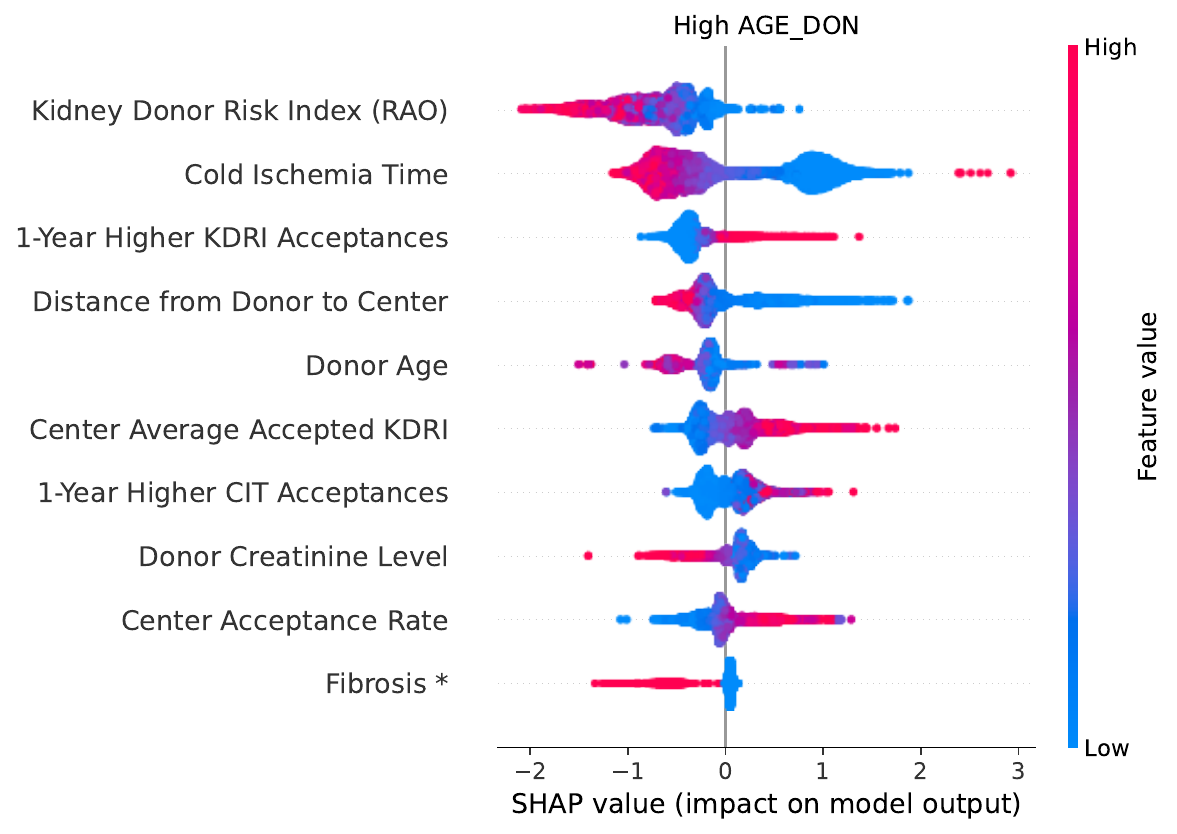}
    \caption{Global Importance Top 10\% of Age}
    \label{fig:enter-label5}
\end{figure}

Figure~\ref{fig:enter-label5} shows that almost all KDRI values in this data subset are lowering acceptance probability. This is likely a result of KDRI accounting for age and the majority of donors in this range have high KDRI values. As a result of the high KDRI values, we see once again that CIT must be significantly lower for an increase in acceptance probability. 

\begin{figure}[!ht]
    \centering
    \includegraphics[width=\linewidth]{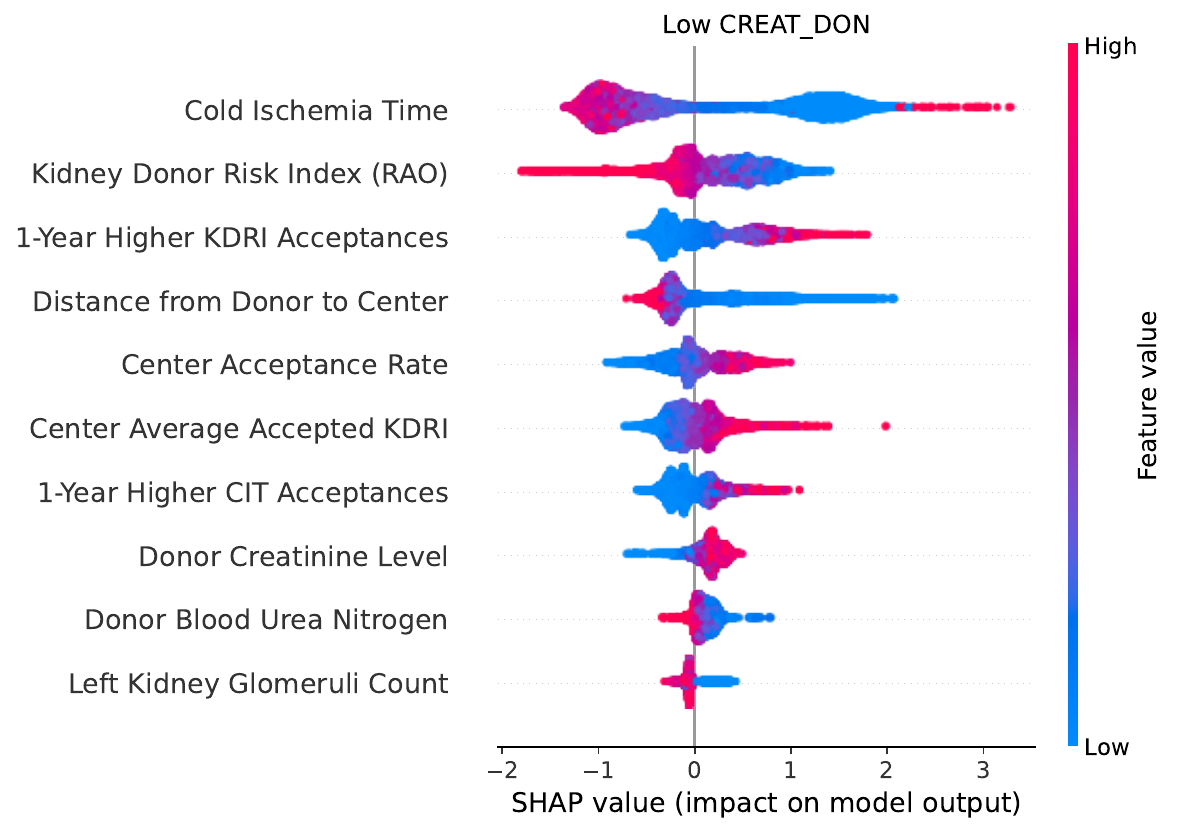}
    \caption{Global Importance Bottom 10\% of Creatinine}
    \label{fig:enter-label6}
\end{figure}

\begin{figure}[!ht]
    \centering
    \includegraphics[width=\linewidth]{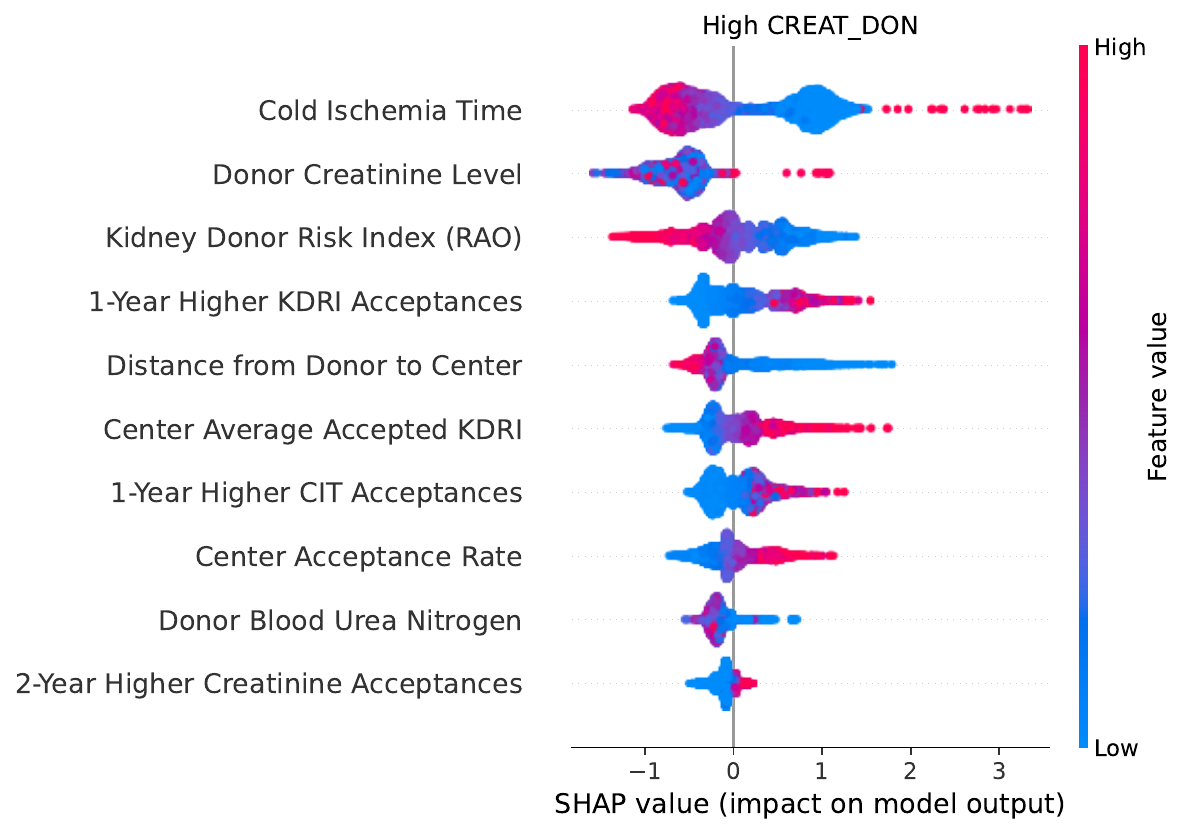}
    \caption{Global Importance Top 10\% of Creatinine}
    \label{fig:enter-label7}
\end{figure}

Figures~\ref{fig:enter-label6} and~\ref{fig:enter-label7} are most interesting when compared together. We note that when the creatinine levels are high which is an indicator of poor kidney function, the impact of CIT is more compact when compared to its low creatinine counterpart. We also note that when the creatinine is very low, the feature itself drops out of the top features altogether. This is because there is not enough variance in their values to impact the model decision. 

\begin{figure}[!ht]
    \centering
    \includegraphics[width=\linewidth]{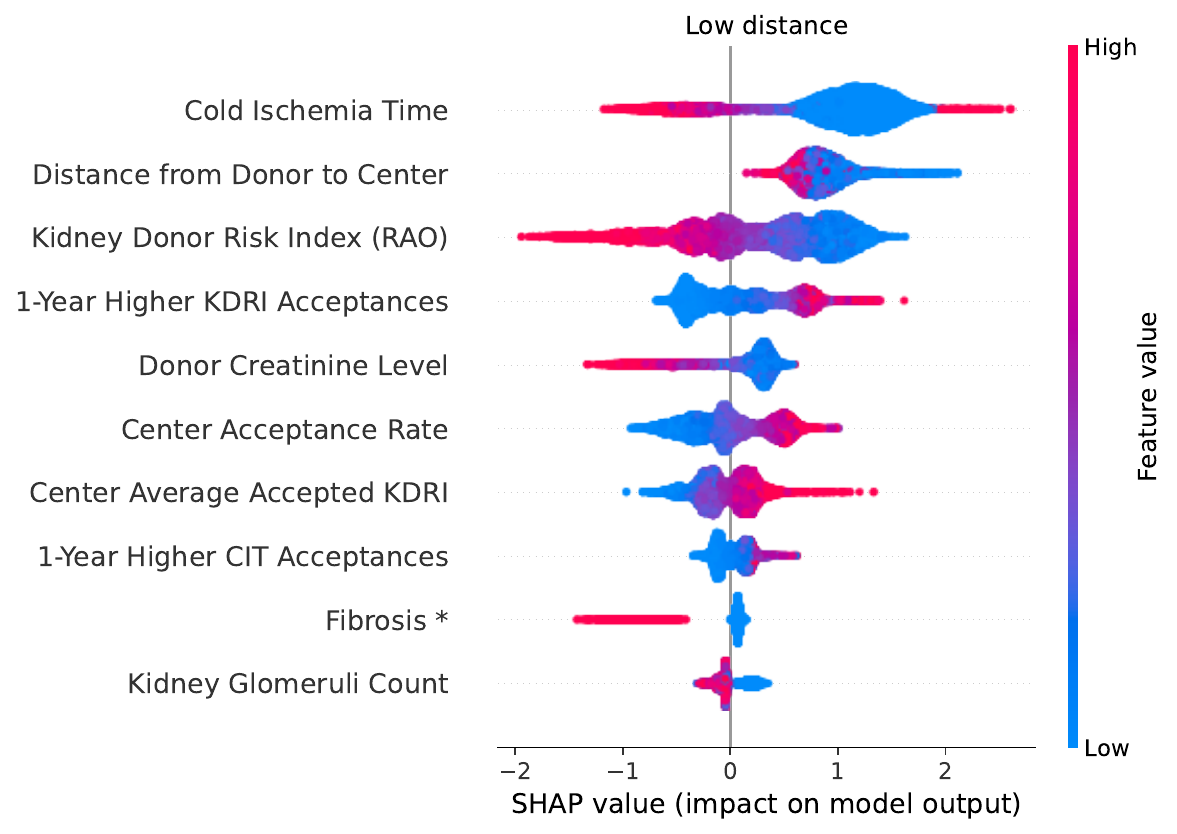}
    \caption{Global Importance Bottom 10\% of Distance}
    \label{fig:enter-label8}
\end{figure}

\begin{figure}[!ht]
    \centering
    \includegraphics[width=\linewidth]{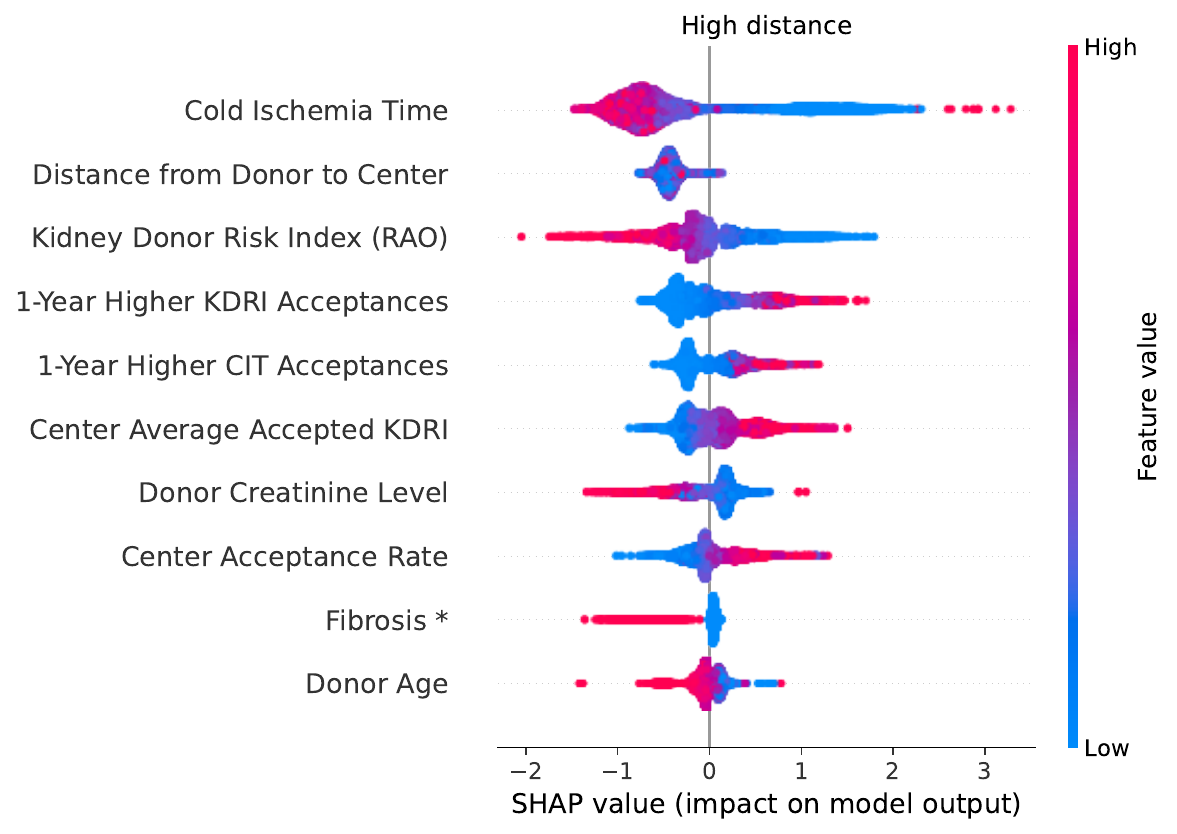}
    \caption{Global Importance Top 10\% of Distance}
    \label{fig:enter-label9}
\end{figure}

Figures~\ref{fig:enter-label8} and~\ref{fig:enter-label9} show the differences in global importance for the top and bottom 10\% of distances. These plots show compelling evidence of the explanatory power of this feature. First, we note that the distance from donor to center becomes the second most important feature for both of these plots. In Figure~\ref{fig:enter-label8}, we see that virtually any donor distance in this range lowers acceptance prediction, whereas in Figure~\ref{fig:enter-label9} shows the opposite i.e., any value is increasing acceptance probability. In addition, CIT appears to be the most important feature in both plots but with significant differences. For the low distances, we see that high CIT is not punished quite as harshly as in the case of the high distance plot. This suggests that centers are willing to accept higher CITs if they can get the organ more quickly. 

%%%%%%%%%%%%%%%%%%%%%%%%%%%%%%%%%%%%%%%%%%%%
\subsection{Local Interpretability Examples}
%%%%%%%%%%%%%%%%%%%%%%%%%%%%%%%%%%%%%%%%%%%%

In Figure~\ref{fig:double_image} we can see that the center-specific features have become much more important when we lack strong donor-specific features. Specifically, we can see that center average accepted KDRI is making up most of the decisions for the top centers. Similarly, we note the center acceptance rate and distance to the donor seem to be important positive features. For the bottom-ranked centers, the patient's fibrosis status seems to be a very slight positive for the acceptance but the center's acceptance rate as well as the patient's CIT seem to make these bad choices.\footnote{Fibrosis Present in Most Portal Areas With Marked Bridging has been shortened to Fibrosis* }

In Figure~\ref{fig:double_image2} we can see that we have more examples of donor-specific features. For the top centers, the positive features seem to be their average accepted KDRI or AGE and the patient's CIT. Also present in the positive features is the donor distance to airports. 
Conversely for the bottom 5 centers, we see almost a reversal the center average acceptances seem to push down acceptance probability significantly. We also see that the ``negative'' features seem to be even more of a problem for these centers e.g., a history of smoking is much more damaging in these centers.

\begin{table}[!ht]
\label{num1}
\centering
\caption{Summary of important features for the case presented in Figure~\ref{fig:double_image}.}
\begin{tabular}{P{0.4\textwidth} P{0.35\textwidth}}
\toprule 
\textbf{Donor Characteristic} & \textbf{Value} \\
\midrule
Donor Blood Type O & False \\
CIT (minutes at first offer) & 594.0 \\
Donor Age (years) & 41.0 \\
Kidney Donor Risk Index & 1.82 \\
Distance to Medium Airport (miles) & 85.72 \\
Distance to Large Airport (miles) & 10.53 \\
Diabetes History: 0-5 years & False \\
Diabetes History: 6-10 years & False \\
Diabetes History: More than 10 years & True \\
Diabetes History: Unknown duration & False \\
Diabetes History: Unsure & False \\
\bottomrule
\end{tabular}
\end{table}

\begin{table}[!ht]
\label{num2}
\centering
\caption{Summary of important features presented in Figure ~\ref{fig:double_image2}.}
\begin{tabular}{P{0.4\textwidth} P{0.35\textwidth}}
\toprule 
\textbf{Donor Characteristic} & \textbf{Value} \\
\midrule
Donor Blood Type O & True \\
CIT (minutes at first offer) & 0.0 \\
Donor Age (years) & 46.0 \\
Kidney Donor Risk Index & 1.71 \\
Distance to Medium Airport (miles) & 5.62 \\
Distance to Large Airport (miles) & 14.15 \\
Diabetes History: 0-5 years & False \\
Diabetes History: 6-10 years & False \\
Diabetes History: More than 10 years & False \\
Diabetes History: Unknown duration & False \\
Diabetes History: Unsure & False \\
\bottomrule
\end{tabular}
\end{table}

\end{document}